\def\ifundefined#1{\expandafter\ifx\csname#1\endcsname\relax}

						\ifundefined{preprint}
\documentclass[fdp,a4paper,fleqn]{w-art}
						\else
						\documentclass[pop,a4paper,
						fleqn]{w-art}	\fi
\usepackage{times,cite}
\usepackage[]{graphicx}
	\numberwithin{equation}{section}
	\numberwithin{figure}{section}

\def\ifundefined#1{\expandafter\ifx\csname#1\endcsname\relax}

\ifundefined{preprint}	\long\def\cut#1\endcut{}  \def\PPnum{}	\def\TALKat{}
\else
	\long\def\cut#1\endcut{#1}
	\def\PPnum{\begin{flushright} \vspace{-2cm}
	{\small NRCPS-HE-56-07 \;TUW-07-15 \;ESI-1992%		,  CERN-???
	}\end{flushright}}
	\def\TALKat{\noindent
	\BP(0,0)\put(299,20){\pspolygon[fillcolor=white,linecolor=white,
	fillstyle=solid](0,0)(150,0)(150,9)(0,9)(0,0)}\EP\\[-23pt]
	Contribution to the proceedings of the BW2007
	Workshop "Challenges Beyond the Standard Model", September 2-9, 2007,
	Kladovo, Serbia	%% BW'2007  To appear in Fort. Phys.
	}
\fi

\usepackage{pst-plot,amssymb}	     \def\mao#1{\mathop{\rm #1}\nolimits} 
\def\BC{\begin{center}}	     \def\EC{\end{center}}
\def\BE {\begin{equation}}      \def\EE {\end{equation}}
\def\BEA{\begin{eqnarray}}      \def\EEA{\end{eqnarray}}
\def\BI{\begin{itemize}}     \def\EI{\end{itemize}}   \def\bye{\end{document}}
\def\BP{\begin{picture}}     \def\EP{\end{picture}}    \let\lila=\magenta
\def\putc#1)#2{\put#1){\makebox(0,0)[c]{#2}}}		\let\nn=\nonumber 
\def\putm#1)#2{\put#1){\makebox(0,0)[c]{$#2$}}}
\def\Pput#1)#2{\BP(0,0)\put#1){#2}\EP} 		\long\def\del#1\enddel{}

\unitlength=1pt \psset{unit=1pt}	\def\kp{\HS-1 \star\HS-1 }

\def\VS#1 {\vspace*{#1pt}}   \def\HS#1 {\hspace*{#1pt}} \let\txt=\textstyle 

\def\cD{{\cal D}}  \def\cF{{\cal F}}  \def\cL{{\cal L}}  
\def\cO{{\cal O}}

\let\bra=\langle   \let\ket=\rangle	\let\then=\Rightarrow  	
\def\2{{1\over2}}  \let\5=\bar	\let\6=\partial	\def\7#1{{#1}\llap{/}}  

\let\a=\alpha   \let\b=\beta    \let\g=\gamma   \let\d=\delta   %% --> GREEK
\let\z=\zeta           \let\e=\varepsilon
\let\k=\kappa   \let\l=\lambda  \let\m=\mu      
\let\n=\nu      \let\x=\xi      \let\p=\pi      \let\r=\rho     \let\s=\sigma 
\let\t=\tau     \let\o=\omega          
\let\Ph=\phi            \let\O=\Omega   \let\S=\Sigma 
      \let\Th=\Theta  \let\L=\Lambda  \let\G=\Gamma   

\let\ns=\normalsize	\let\fns=\footnotesize

   \newcounter{TRefNX}   \makeatletter%    DRAFT MODE macros
   \def\makeTRefs#1{\@for  \NewTRef:=#1\do{\global\makeTRef{\NewTRef}}}
   \def\makeTRef#1{\ifundefined{TRef#1}\stepcounter{TRefNX}%
   \expandafter\xdef\csname TRef#1\endcsname{\theTRefNX}\fi}\makeatother
   \def\LLab#1{\BP(0,0)\unitlength=1mm\put(-9,-1.6){\makebox(0,0)[cr]{\small#1
        \rlap{$_{_{\makeatletter\csname TRef#1\endcsname\makeatother}}$}}}\EP}
   \def\BP{\begin{picture}} \def\EP{\end{picture}} \let\OLDbib=\bibitem 
	\def\NEWbib#1{\OLDbib{#1}\LLab{#1}}		
\ifundefined{draftmode} \else 
	\let\bibitem=\NEWbib \fi

\newcounter{Bem} % this line was added by Sebastian at November 27,2007

\newcommand{\bs}[1]{\boldsymbol{#1}}

\newcommand{\mc}[1]{\mathcal{#1}}

\newcommand{\bei}[2]{\left. #1 \right| _{#2 }}

\newcommand{\q}[1]{\underline{#1 }}
\newcommand{\hoch}[1]{{}^{#1}}

\newcommand{\os}[2]{\overset{\lefteqn{{\scriptstyle #1}}}{#2}}

\newcommand{\ce}{\bs{\lambda}}
\newcommand{\be}{\bs{\omega}}
\newcommand{\RR}{{\mathbb P}}
\newcommand{\rr}{P}

\newcommand{\dP}{d}

\newcommand{\xfull}{\,\os{{\scriptscriptstyle \twoheadrightarrow}}{x}\,}
\newcommand{\xboson}{\,\os{{\scriptscriptstyle \rightarrow}}{x}\,}

\begin{document}						\PPnum

%%    The information for the title page will be placed between
%%    \begin{document} and \maketitle. The order of most entries
%%    is determined by the class file and can not be changed by
%%    rearranging them. The maketitle command follows after the
%%    abstract.
%%
%%    Most of the following commands will be completed by the publisher.
%%
%%    The copyrightyear is defined in the .clo file as the first argument
%%    of the copyrightinfo command. If the copyrightyear differs from that
%%    value it might be adjusted by the following definition:
%%
%% \renewcommand{\copyrightyear}{2007}% uncomment to change the copyrightyear.
%%
%--%	\DOIsuffix{theDOIsuffix}
%%
%% issueinfo for the header line
%--%	\Volume{55}
%--%	\Month{01}
%--%	\Year{2007}
%%
%%    First and last pagenumber of the article. If the option
%%    'autolastpage' is set (default) the second argument may be left empty.
\pagespan{1}{}
%%
%%    Dates will be filled in by the publisher. The 'reviseddate' and
%%    'dateposted' (Published online) entry may be left empty.
%--%	\Receiveddate{XXXX}
%--%	\Reviseddate{XXXX}
%--%	\Accepteddate{XXXX}
%--%	\Dateposted{XXXX}
%%

\keywords{Noncommutative geometry, D-branes, deformation quantization}
\subjclass[pacs]{02.40.Gh % Noncommutative geometry 
	%%	04.60.-m Quantum gravity
	%%	04.60.Ds Canonical quantization 
		04.62.+v % Quantum field theory in curved spacetime 
		11.25.-w % Strings and branes
		11.25.Uv % D branes
				%--%	04A25%%PACS-Numbers
%\qquad\parbox[t][2.2\baselineskip][t]{100mm}{%
%  \raggedright
%  (Please use PACS-codes from the enclosed list
%  (ASCII2006FullPACS.txt) or from www.aip.org/pacs)\vfill}
}%

%% \pretitle{Editor's Choice}

%% We have a short and a long form for the title. The short form
%% (optional argument) goes into the running head.

\title[Non-topological non-commutativity]
	{Non-topological non-commutativity in string theory}

%% Please do not enter footnotes or \inst{}-notes into the optional
%% argument of the author command. The optional argument will go into
%% the header.  If there is only one address the marker \inst{x} may be
%% omitted.

%% Information for the first author.
\author[S. Guttenberg]{Sebastian Guttenberg\inst{1,}%
  \footnote{\textsf{guttenb@inp.demokritos.gr}
%--%	    Corresponding author\quad E-mail:~\textsf{x.y@xxx.yyy.zz},
%--%            Phone: +00\,999\,999\,999,
%--%            Fax: +00\,999\,999\,999
}}
\address[\inst{1}]{NCSR Demokritos, INP, Patriarchou Gregoriou \& 
	Neapoleos Str., 15310 Agia Paraskevi Attikis, GREECE}
%%
%%    Information for the second author
\author[M. Herbst]{Manfred Herbst\inst{2,}\footnote
			{~ \textsf{manfred.herbst@cern.ch}}}
\address[\inst{2}]{CERN, 1211 Geneva 23, SWITZERLAND}
%%
%% Information for the third author.
\author[M. Kreuzer]{Maximilian Kreuzer\inst{3,}%
  \footnote{%E-mail:~\textsf{maximilian.kreuzer@tuwien.ac.at}%,
	\textsf{maximilian.kreuzer@tuwien.ac.at}
%            Phone: +00\,999\,999\,999,
%            Fax: +00\,999\,999\,999
}}
\address[\inst{3}]{Institute for Theoretical Physics, TU--Wien, 
	Wiedner Hauptstr. 8--10, 1040 Vienna, AUSTRIA}
%%
%%    Information for the fourth author
\author[R. Rashkov]{Radoslav Rashkov\inst{4,}\footnote
	{~ ~\textsf{rash@hep.itp.tuwien.ac.at}}}
\address[\inst{4}]{Erwin Schr\"odinger Institute for Mathematical Physics, 
	Boltzmanngasse 9, 1090 Vienna, AUSTRIA}
%%
%%    \dedicatory{This is a dedicatory.}
\begin{abstract}
Quantization of coordinates leads to the non-commutative product of 
deformation quantization, but is also at the roots of string theory, for which
space-time coordinates become the dynamical fields of a two-dimensional
conformal quantum field theory. Appositely, open string 
diagrams provided the inspiration for Kontsevich's solution of the 
long-standing problem of quantization of Poisson geometry by virtue of his 
formality theorem. 
%% The emergence of the corresponding formulas from a 
%% topological limit of open strings was then shown by Cattaneo and Felder. 
In the context of D-brane physics non-commutativity is not limited, however, 
to the topolocial sector. We show that non-commutative effective actions 
still make sense when associativity is lost and establish a generalized 
Connes-Flato-Sternheimer condition %, termed cyclicity by Felder and Shoikhet,
through second order in a derivative expansion. The measure in 
general curved backgrounds is naturally provided by the Born--Infeld action 
and reduces to the symplectic measure in the topological limit, but 
remains non-singular even for degenerate Poisson structures. 
%% Its compatibility condition with the deformed product is equivalent to the
%% generalized Maxwell equation for the background gauge field. 
Analogous superspace deformations by RR--fields are also discussed. 
\end{abstract}
%% maketitle must follow the abstract.
\maketitle                   % Produces the title.

\TALKat

\section{Introduction}

The non-commutative product of deformation quantization 
	\cite{Kontsevich:1997vb,Sternheimer:1998yg} can 
be derived from string theory in a topological limit where the space-time 
metric is small as compared to the anti-symmetric B-field (the ancestor of 
the Poisson bi-vector) \cite{Schomerus:1999ug,Seiberg:1999vs,Cattaneo:2000fm}.
The non-commutative
product thus amounts to a summation of the leading B-field contributions
to the effective action. In the non-symplectic case this interpretation 
is spoiled, however, by the absence of a canonical measure. From the string
theory point of view, on the other hand, associativity is lost for generic
backgrounds \cite{Cornalba:2001sm}, but the 
Born-Infeld action provides a canonical measure 
\cite{Seiberg:1999vs,Herbst:2001ai}. We show that the concept of effective 
actions does not require associativity, but rather a generalized
Connes--Flato--Sternheimer condition called cyclicity 
\cite{Schoikhet:1999,Felder:2000nc}, i.e. commutativity and associativity 
up to surface terms\cite{cyclic}. Cyclicity implies, however,
a compatibility condition between the star product and the
measure \cite{Felder:2000nc}, which for Born-Infeld turns out to be equivalent
to the generalized Maxwell equation for the gauge field on the D-brane 
\cite{Herbst:2001ai,cyclic}. In \cite{cyclic} we found that cyclicity also
requires a gauge modification of the Kontsevich product at second derivative 
order in a derivative expansion and we discussed the D-brane physics related 
to these mathematical structures.

In section 2 of this note we review some aspects of deformation quantization 
and formality
in simple terms by illustrating the emergence of Hochschild cocycles, 
Gerstenhaber brackets and gauge transformations accompanying diffeomorphisms 
in derviative expansions.
In section 3 we discuss the stringy origin of these structures and their
interpretation in terms of effective actions, which requires
the existence of a measure and a generalized Connes--Flato--Sternheimer
property. While associativity is restricted to 
Poisson geometry, string theory naturally introduces the Born--Infeld measure
and keeps cyclicity, at least through second derivative order, independently
of associativity and without a topological limit.  We observe that
the results found in \cite{cyclic} straightforwardly extend to non-constant
dilaton backgrounds. In section 5 we discuss the 
Berkovits string in general RR backgrounds and the resulting deformation of
superspace. In section 6
we conclude with a discussion of open problems and work to be done.

\section{Deformation quantization, Kontsevich product and formality}

The idea of deformation quantization is to emulate the operator product of
quantum mechanics by an associative product $f\kp g$ of phase
space functions $f,g\in C^\infty(M)$ with 		\VS-5
\BE						\label{PBstar}
	f\kp g=f\,g+\frac i2\hbar\{f,g\}_{PB}+\cO(\hbar^2)\quad\then\quad
	\lim_{\hbar\to0}\;\frac{f\kp g-g\kp f}{i\hbar}=\{f,g\}_{PB},
\EE				\\[-12pt]
where the Poisson bracket can be written for arbitrary phase space coordinates 
$x^\m$ as a bi-derivation
$
	 \{f,g\}_{PB}=\Th^{\m\n}(x)\6_\m f\6_\n g	
$
in terms of a bi-vector field $\Th\in\L^2 TM$. 

\subsection{Polyvectors and the Schouten--Nijenhuis bracket}

Elements $X\in\L^\bullet TM$ of the exterior algebra over the tangent 
space $TM$ are called polyvector fields and there is a bilinear operation, 
the Schouthen--Nijenhuis (SN) bracket		\VS-3
\BE
	[X^{(p)},Y^{(q)}]\in\L^{p+q-1}TM\quad \hbox{for}\quad
		X^{(p)}\in \L^{p}TM \quad\hbox{and}\quad Y^{(q)}\in \L^{q}TM,
\EE					\\[-12pt]
that extends the Lie bracket of vector fields to a
graded bi-derivation of degree $-1$ on $T\in\L^\bullet TM$. 
The Jacobi identity of the Poisson bracket is 
equivalent to the vanishing of the SN bracket $[\Th,\Th]$,	\VS-3
\BE	\thicklines \unitlength=1pt	
	\BP(0,0)\put(7.6,2.9){\circle6}\EP\sum \{\{f,g\}_{PB},h\}_{PB}=0
	\quad\Leftrightarrow\quad [\Th,\Th]=0 \quad\hbox{with}
	\quad [\Th,\Th]^{\m\a\b}=\frac23\BP(0,0)\put(7.6,2.9){\circle6}\EP
		\sum\Th^{\m\r}\6_\r\Th^{\a\b}.
\EE					\\[-12pt]
Lie derivatives in the direction of $\x\in TM$ can 
also be written in terms of the SN bracket
$				%		\label{LieDeriv}
	\cL_\x X=[\x,X]		%\quad\hbox{for}\quad	X\in \L^\bullet TM.
$
for all polyvector fields $X\in \L^\bullet TM$.

\subsection{Moyal product and Kontsevich graphs}

In case of constant $\Th$, and hence in particular locally for Darboux 
coordinates, deformation quantization can be achieved by the Moyal product
							\VS-3
\BE
	% f(x)\cdot g(x)~\to~
	(f\kp g)(x)=\mao{exp}\left(\txt\frac i2\hbar\,
		\Th^{\m\n}\6_{y^\m}\6_{z^\n}\right)
		f(y)g(z)\;_{{\,\vrule height 3mm depth 2pt width .4pt}
		_{\,\hbox{\fns$y=z=x$}}}
\EE							\del
\\[-12pt]which is associative since		\VS-3
\BEA
	((f\kp g)\kp h)(x)&=&e^{\frac {i\hbar}2\Th^{\m\n}(\6_{u^\m}+\6_{v^\m})
		\6_{w^\n}}e^{\frac{i\hbar}2\Th^{\a\b}{\6_{u^\a}\6_{v^\b}}}
		f(u)g(v)h(w)\;_{{\,\vrule height 3mm depth 2pt width .4pt}
		_{\,\hbox{\fns$u=v=w=x$}}}	
\\	&=&e^{\frac{i\hbar}2\Th^{\a\b}(\6_{u^\a}\6_{v^\b}+\6_{u^\a}\6_{w^\b}
		+\6_{v^\a}\6_{w^\b})}f(u)g(v)h(w)\;_{{\,
	   \vrule height 3mm depth 2pt width .4pt}_{\,\hbox{\fns$u=v=w=x$}}}
\\	&=&(f\kp (g\kp h))(x)
\EEA
\\[-17pt]for constant $\Th$.

\begin{figure}
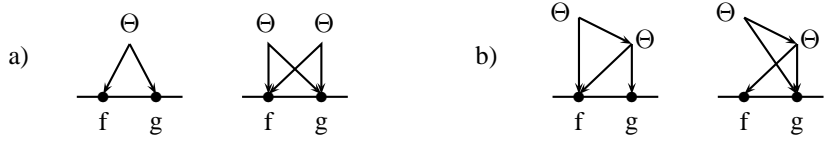

\BP(129,50)(-55,-12)	%\blue\psset{linecolor=blue}
\put(0,0){   \psline(0,0)(40,0)\put(10,0){\circle*4}\put(30,0){\circle*4}
	\putc(10,-9)f\putc(30,-11)g	\putc(-22,15){a)}
   \psline{->}(20,20)(30,1)\psline{->}(20,20)(10,1)\putc(20,28){$\Th$}

\put(60,0){
   \psline(0,0)(40,0)\put(10,0){\circle*4}\put(30,0){\circle*4}
	\putc(10,-9)f\putc(30,-11)g
   \psline{<->}(10,1)(10,20)(30,1)\putc(9,28){$\Th$}\putc(31,28){$\Th$}
   \psline{<->}(10,1)(30,20)(30,1)}}

\put(180,0){{		%\lila\psset{linecolor=magenta}
   \psline(0,0)(40,0)\put(10,0){\circle*4}\put(30,0){\circle*4}
	\putc(10,-9)f\putc(30,-11)g	\putc(3,32){$\Th$}
   \psline{->}(10,30)(10,1)\psline{->}(10,30)(30,20)	\putc(35,22){$\Th$}
   \psline{->}(30,20)(10,1)\psline{->}(30,20)(30,1)	\putc(-25,15){b)}}

\put(60,0){		%\lila\psset{linecolor=magenta}
   \psline(0,0)(40,0)\put(10,0){\circle*4}\put(30,0){\circle*4}
	\putc(10,-9)f\putc(30,-11)g	\putc(3,32){$\Th$}
   \psline{->}(10,30)(30,1)\psline{->}(10,30)(30,20)	\putc(35,22){$\Th$}
   \psline{->}(30,20)(10,1)\psline{->}(30,20)(30,1)		}}
\EP		%\black\psset{linecolor=black}
\caption{Kontsevich graphs for a) Moyal-type contributions and b) derivative 
	corrections, respectively.}	\label{formality}	\VS-12
\end{figure}

							\enddel
After a general change of coordinates in phase space $\Th$ will not stay 
constant, which motivates the consideration of deformation quantization 
for general $\Th$. For the symplectic case $\det\Th\neq0$ 
the existence of a star product has been shown by De Wilde and Lecompte 
\cite{DeWiLe83} and the first construction is due to Fedosov \cite{Fedosov}.
Some details and a historical assessment with references can be found in the 
review \cite{Sternheimer:1998yg}.
For the case of a general Poisson structure $\Th$, which by definition 
obeys $[\Th,\Th]=0$, the construction of an associative product is due 
to Kontsevich \cite{Kontsevich:1997vb} and will now be discussed in more 
detail. Associativity of this product is, in fact, a corollary of the 
formality theorem, which establishes a quasi-isomorphism of $L_\infty$ 
algebras. The formality map $U$ maps polyvector fields $T_i$ to 
polydifferential operators 
%				\cut	\VS-3	\endcut
$ %\BE						\label{Fmap}
	U(T_1,\ldots,T_n)=\sum_\G w_\G D_\G
$ %\EE			%% Kontsevich [q-alg/9709040]
%							\\[-15pt]
and is constructed in terms of graphs $\G$ and coefficients $w_\G$. The
coefficients $w_\G$ are defined by convergent integrals 
inspired by open string Feynman diagrams (cf. section 3)
with functions inserted on the real line and polyvector fields in the
upper half plane as illustrated in fig.\,\ref{formality}, where %Here 
the derivatives of the bidifferential operators
correspond to the arrows pointing at $f$ and $g$. The first two graphs
fig. \ref{formality}a give the order $\Th$ and $\Th^2$ terms of the Moyal 
product, while fig.\,\ref{formality}b yields first derivative corrections
for non-constant $\Th$. The latter will be worked out explicitly below.
The precise relation of Kontsevich's construction to correlation 
functions of topological sigma models is due to
Cattaneo and Felder \cite{Cattaneo:2000fm}.

\del
	\blue Poisson case: \black Kontsevich [q-alg/9709040]
\largE
\smallskip
\BI\item	\lila \bf Formality map~\rm 
		\blue $U$: Polyvector $T$ $\to$ Poly-DifferentialOperator $D$	\black
\smallskip
      \item $U(T_1,\ldots,T_n)=\sum_\G w_\G D_\G$ \quad $w_\G=$
	convergent integrals \ns(string-inspired)\largE
\item	Cattaneo--Felder: = topological $\s$-model perturbative expansion
 \item \lila Quasi-isomorphism of $\cL_\infty$ algebras \black
\\	Hochschild complex (cf. Ivo's)
	%dg Lie algebras (\bf d\rm ifferential: $Q^2=0$; \bf g\rm raded) 
\\	\blue Schouten--Nijenhuis $\mapsto$ Gerstenhaber bracket 
	\black	\hbox{\ns(=``commutator'' of PDOs)}		    \EI	\EI

\enddel

\subsection{Hochschild cohomology, Gerstenhaber bracket, and the formality
	theorem}

Rather than giving abstract definitions of the involved mathematical 
structures we now illustrate how they automatically show up in simple
calculations. We ignore for a moment the relation (\ref{PBstar}) 
to Poisson brackets and consider a general deformation of the product\VS-4
\BE					\label{DefoHbar}
	f\kp g=fg+\hbar B_1(f,g)+\cO(\hbar^2)\qquad \hbox{with}\qquad 
		B_1(f,g)=B^{\m\n}f_\m g_\n, \quad f_\m\equiv\6_{x^\m}f,
\EE						\\[-14pt]
where derivatives of functions are abbreviated by subscripts. The 
$\cO(\hbar)$ contribution to the associator,				\VS-4
\BE
	f\kp(g\kp h)-(f\kp g)\kp h
	=\hbar\Bigl( fB_1(g,h)-B_1(fg,h)+
		B_1(f,gh)
		-B_1(f,g)h	\Bigr)+\cO(\hbar^2),
\EE					\\[-14pt]
has exactly the form of a Hochschild coboundary \cite{Kontsevich:1997vb}\VS-4
\BE
	(\d C)(f_0,\ldots,f_p)=f_0C(f_1,\ldots,f_p)-C(f_0f_1,\ldots,f_p)
	+C(f_0,f_1f_2,\ldots,f_p)-\ldots
\EE							\\[-12pt]
There are, however, equivalences of the resulting deformed associative
algebras due to invertible maps $f\to Df$ with differential operators\VS-4
\BE						\label{GaugeEq}
%	f\to Df\qquad\hbox{with}\qquad 
	D=1+\hbar(D_1^\m\6_\m+D_1^{\m\n}\6_\m\6_\n+\ldots)
	+\hbar^2(D_2^\m\6_\m+\ldots)+\ldots
\EE								\\[-12pt]
that respect the unit element $D1=1$. They lead to the following
modification of the star product,				\VS-4
\BE						\label{EquiTrans}
	f\to Df \qquad\then\qquad f\kp'g=D(D^{-1}f\,\kp\,D^ {-1}g)
\EE							\\[-14pt]
and hence 				
$
	B'_1(f,g)-B_1(f,g)=-fD_1(g)+D_1(fg)-D_1(f)g,
$ at order $\hbar$, which is again a Hochschild coboundary. 
For the special case $D_1=D_1^{\m\n}\6_\m\6_\n$ this implies the gauge 
equivalence $B'_1(f,g)-B_1(f,g)=D_1^{\m\n}f_\m g_\n$ so that 
for the first order bidifferential operator $B_1(f,g)=B_1^{\m\n}f_\m g_\n$
of eq. (\ref{DefoHbar})\,
the symmetric part of $B_1^{\m\n}$ can be gauged away with 
$D_1^{\m\n}=B_1^{(\m\n)}$. With the choice 
$B_1^{\m\n}=\frac i2\Th^{\m\n}$ we thus recover (\ref{PBstar}).

Returning to the Kontsevich graphs fig. \ref{formality} we now want to
work out the derivative corrections that are needed for associativity
at order $\hbar^2$. For this purpose we define the Moyal part 
%% $[f\star g]$ of a product $f\star g$						\VS-4
\BE	\txt
	[f\kp g]
	\equiv fg+i\frac \hbar2\Th^{\m\n} f_\m g_\n
	-\frac{\hbar^2}8\Th^{\m\a}\Th^{\n\b}f_{\m\a}g_{\n\b}-\ldots
\EE								\\[-12pt]
% of a product, 
of a product $f\star g$	as the result of dropping all terms with 
derivatives acting on $\Th$. %indicated by square brackets $[]$. 
Then				\VS-4
\BE		
	f\kp g= [f\kp g]-\hbar^2\,\Th^{\m\r}\6_\r\Th^{\a\b}(a\,
	   [f_{\a\m} \kp g_\b]+b\,[f_\a \kp g_{\b\m}])\black
	+\cO\bigl(\6^2\bigr)
\EE						\\[-12pt]
where $\cO\bigl(\6^2\bigr)$ only counts derivatives acting on $\Th$ and
the coefficients $\o_\G$ of the two graphs in fig. \ref{formality}b
are $a$ and $b$, respectively. Instead of determining these coefficients
from integrals over $\Th$ in the upper half plane we determine them by
imposing associativity.
The first derivative order 
part of $f\kp(g\kp h)$ is
%
%
%With $X^{\m\a\b}=\Th^{\m\r}\6_\r\Th^{\a\b}$ the first derivative order 
%part %% $\cO(\6)$ part 
% of $f\kp(g\kp h)\!=\![f\kp g\kp h]$ is
		\VS-5
\BE									\VS-5
	%\txt\!\!\!\nonumber f\kp(g\kp h)\!=\![f\kp g\kp h]\!
\HS-25	-\hbar^2 X^{\m\a\b}\Bigl[a f_{\m\a}\kp			
		(g\kp h)_\b\!+\! b f_{\a}\kp(g\kp h)_{\m\b}\!	
		+\!\frac14f_\m\kp(g_\a\kp h_\b)		
%\VS-3 \EE\BE\txt\HS134	
	+af\kp g_{\m\a}\kp h_\b+ b f\kp g_\a \kp h_{\m\b})\Bigr]	
\EE						
with $X^{\m\a\b}=\Th^{\m\r}\6_\r\Th^{\a\b}$, and the $\cO(\6)$ contributions
to $(f\kp g)\kp h$ are							\VS-5
\BE	%\txt\!\!\!(\!f\kp g)\kp h\!=\![f\kp g\kp h]\!
\HS-25	-\hbar^2 X^{\m\a\b}
		\bigl[a(f\kp g)_{\m\a}\kp h_\b\!
	+\!b\,(f\kp g)_{\a}\kp h_{\m\b}\!-\!\frac14(f_\a\kp g_\b)\kp h_\m
%\VS-3 \EE\BE\txt\HS134	
	+a f_{\m\a}\kp g_{\b}\kp h+ b f_\a\kp g_{\m\b} \kp h)\bigr]
\EE							\\[-12pt]
so that 
% Collecting all first derivative terms and using the antisymmetry of $\Th$
\del\BE	\txt
	%	[\,\kp,\kp\,](f,g,h)=
	f\,\kp\,(g\,\kp \,h)\!-\!(\!f\,\kp\, g)\kp h=X^{\m\a\b}\!\left(
		(a\!-\!\frac14)f_\m g_\a h_\b\!-\!(a\!-\!b)f_\b g_\m h_\a
		+(b\!-\!\frac14)f_\a g_\b h_\m\right)
\EE
\enddel
$	\txt
	%	[\,\kp,\kp\,](f,g,h)=
	f\,\kp\,(g\,\kp \,h)\!-\!(\!f\,\kp\, g)\kp h=\hbar^2 [f_\m*g_\a*h_\b]
	\Bigl(
		(a\!-\!\frac14)X^{\m\a\b} \!+\!(a\!-\!b)X^{\a\m\b}
		-(b\!+\!\frac14)X^{\b\m\a}\Bigr).
$
Associativity implies that the coefficient of $[f_\m*g_\a*h_\b]$ vanishes.
Using the antisymmetry of $\Th$ we first observe that
$X$ cannot be totally antisymmetric. Thus symmetrization in $\m\a$, $\a\b$
and $\b\m$ implies $b=2a-\frac14$, \,$a=2b+\frac14$ and $a+b=0$, respectively.
The unique solution is $a=-b=\frac1{12}$. Hence		\VS-4
\BE
	f\,\kp\,(g\,\kp \,h)\!-\!(\!f\,\kp\, g)\kp h=-\frac16\hbar^2 
		[f_\m*g_\a*h_\b]\BP(0,0)\put(7.6,2.9){\circle6}\EP
	\sum_{\m\a\b}\Th^{\m\r}\6_\r\Th^{\a\b}+\cO(\6^2)	\label{assoc}
\EE						\\[-12pt]
so that $[\Th,\Th]=0$ is a necessary condition for the existence of an
associative deformation. The Kontsevich product through second derivative 
order (setting $\hbar=1$)				\VS-4
		\newcommand {\Back}{\!\!\!\!\!}\newcommand {\Ii} {\mathrm{i}}
\begin{eqnarray}
 \label{eq:finproduct} 
 \HS-25	f \kp g &\HS-12 =\HS-12 & [f \kp g] 
    -\frac{1}{12}\Th^{\mu\gamma}\partial_{\gamma}\Th^{\nu\rho}
       \bigl[f_{\mu\nu}\kp g_{\rho}+f_{\rho}\kp g_{\mu\nu}\bigr] \nonumber 
       + \frac{1}{24} \partial_\sigma \Th^{\mu\rho}
                         \partial_\rho   \Th^{\nu\sigma}
                         [f_\mu \kp g_\nu]
\\
\HS-12 &&\HS-12 +\,\frac{\Ii}{48}\Th^{\mu\gamma}\6_{\gamma}\Th^{\nu\delta}
  \partial_{\delta}\Th^{\rho\lambda}~
  [\,f_{\mu\rho}\kp g_{\nu\lambda}
   \!-\!f_{\nu\lambda}\kp g_{\mu\rho}] \nonumber
 \!-\!\frac{\Ii}{48}\Th^{\mu\gamma}\Th^{\nu\delta}
     \partial_{\gamma}\partial_{\delta}\Th^{\rho\lambda}~
     \bigl[f_{\m\n\r}\kp g_{\lambda}\!-\!f_{\lambda}\kp g_{\mu\nu\rho}\bigr] 
     \nonumber 
\\
\HS-12 &&\HS-12 +\,\frac{1}{2}\frac{1}{12^2}
      (\Th^{\mu\gamma}\partial_{\gamma}\Th^{\rho\lambda})
      (\Th^{\nu\delta}\partial_{\delta}\Th^{\sigma\tau})
      \bigl[f_{\mu\rho\nu\sigma}\kp g_{\lambda\tau} 
       \!+\!2 f_{\mu\rho\tau}\kp g_{\lambda\nu\sigma}
       \!+\! f_{\lambda\tau}\kp g_{\mu\rho\nu\sigma}\bigr]
	+\cO(\6^3)						\label{KP}
\end{eqnarray}				\\[-12pt]
has been determined in \cite{cyclic} using the known coefficient $\frac1{24}$
of the gauge term and symmetry under complex conjugation combined with the
exchange of $f$ and $g$. Note that each term in (\ref{KP}) comprises the
contributions of an infinite number of graphs with Moyal-type additions to 
the classical part of $\kp$, since this formula holds to all orders in 
the undifferentiated $\Th$'s.

The Gerstenhaber bracket of polydifferential operators $P_i$ is the 
commutator $[P_1,P_2]$ with respect to an appropriate definition of the
composition $P_1\circ P_2$ of degree $-1$. For bidifferential operators
the bracket thus yields a tridifferential operator, and in the special case
$P_1=P_2=\kp$\, the bracket becomes proportional to the associator
$\frac12[\kp,\kp](f,g,h)=f\kp(g\kp h)-(f\kp g)\kp h$. The formality map 
% \cut (\ref{Fmap})\endcut 
can be regarded as a dressing, or quantization, of 
polyderivations $T_i\in \L^\bullet TM$ to higher order polydifferential
operators $P_i$. The formality theorem ensures that this map is an 
$\cL_\infty$ quasi-isomorphism where the (homotopy) Lie algebra structures 
are related to the SN bracket and the Gerstenhaber bracket, respectively.

\begin{figure}
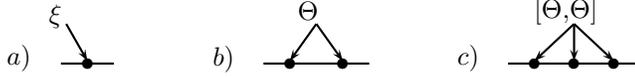

\BP(150,30)(-53,-3)
\put(1,0){	\psline(0,0)(20,0)\put(10,0){\circle*4}	\putc(-16,2){$a)$}
	\psline{->}(2,15)(10,1)\putc(-2,18){$\x$} }
\put(80,0){		\putc(-15,2){$b)$}	%\black\psset{linecolor=black}
	\psline(0,0)(40,0)\put(10,0){\circle*4}\put(30,0){\circle*4}
      \psline{->}(20,15)(10,1)\psline{->}(20,15)(30,1)\putc(17,20){$\Th$} 
}
\put(170,0){	%\red\psset{linecolor=red}
	\psline(0,0)(50,0)\put(10,0){\circle*4}\put(25,0){\circle*4}
	\put(40,0){\circle*4}
      \psline{->}(25,15)(10,1)\psline{->}(25,15)(25,1)\psline{->}(25,15)(40,1)
	\putc(22,20){$[\Th,\!\Th]$} 			\putc(-15,2){$c)$}
}
    \EP
\caption{Dressings of a) Lie derivative, b) Poisson bracket and c) associator,
	respectively.}				\VS-12	\label{fig123}
\end{figure}

The cases of vector fields $\x$, Poisson tensors $\Th$ and rank three 
tensors $J\in\L^3 TM$ shown in fig.\,\ref{fig123} are of particular 
interest. The quantization of $\Th$ yields the star product (\ref{KP}).
Since the SN bracket is mapped to the Gerstenhaber bracket, $J=[\Th,\Th]$
as well as its quantization vanish in the case of a Poisson structure
$[\Th,\Th]=0$. Since $[\kp,\kp]$ is the associator this establishes 
associativity of the Kontsevich product (compare fig.\,\ref{fig123}c
to our result (\ref{assoc}) at leading order $\hbar^2$).

\begin{figure}	[h]
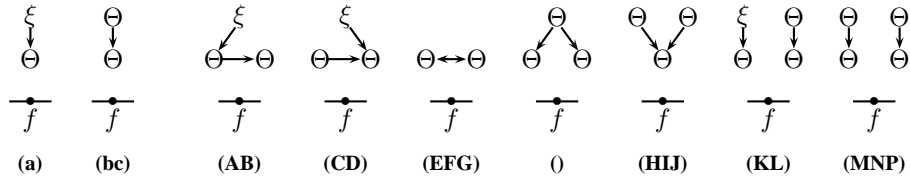
	\unitlength=.8pt	\psset{unit=\unitlength}
\BC
\BP(180,70)(20,-22)	
	\def\putlcf{\psline(-10,0)(10,0)\put(0,0){\circle*4}\putm(0,-9)f}
	\def\putlcf{\psline(-10,0)(10,0)\put(0,0){\circle*4}\putm(0,-9)f}

%\put(-100,0){	\putlcf  			\putc(0,-30){\fns\bf(a)}
%	 \putm(-12,20)\x \psline{->}(-9,20)(6,20) \putm(12,20)\Th}
\put(-99,0){	\putlcf				\putc(0,-30){\fns\bf(a)}
	\putm(0,20)\Th 	\putm(0,40){\x}	\psline{->}(0,35)(0,25)}
\put(-60,0){	\putlcf				\putc(0,-30){\fns\bf(bc)}
	\putm(0,20)\Th 	\putm(0,40)\Th	\psline{->}(0,35)(0,25)}
%\put(100,0){  \psline(-10,0)(10,0)\put(0,0){\circle*4}
%	\putm(-12,20)\x	\psline{<-}(-9,20)(7,20) \putm(12,20)\Th 
%	\putm(0,40)\Th	\psline{->}(3,35)(9,24)	\putc(0,-30){\fns\bf(1c)} }

\put(00,0){	\putlcf				\putc(0,-30){\fns\bf(AB)}
	\putm(-12,20)\Th\psline{->}(-9,20)(7,20)\putm(12,20)\Th 
	\putm(0,40)\x \psline{->}(-2,35)(-10,24)}
\put(50,0){	\putlcf				\putc(0,-30){\fns\bf(CD)}
	\putm(-12,20)\Th\psline{->}(-9,20)(7,20)\putm(12,20)\Th 
	\putm(0,40)\x \psline{->}(2,35)(10,24)}
\put(100,0){	\putlcf				\putc(0,-30){\fns\bf(EFG)}
	\putm(-12,20)\Th\psline{<->}(-7,20)(7,20)\putm(12,20)\Th }
\put(150,0){	\putlcf				\putc(0,-30){\fns\bf()}
	\putm(-12,20)\Th\psline{->}(-2,35)(-10,24)\putm(12,20)\Th 
	\putm(0,40)\Th \psline{->}(2,35)(10,24)}
\put(200,0){	\putlcf				\putc(0,-30){\fns\bf(HIJ)}
	\putm(-12,40)\Th\psline{<-}(-2,24)(-10,35)\putm(12,40)\Th 
	\putm(0,20)\Th \psline{<-}(2,24)(10,35)}
\put(250,0){	\putlcf				\putc(0,-30){\fns\bf(KL)}
	\putm(-12,20)\Th \putm(12,20)\Th \psline{->}(12,35)(12,25)
	\putm(12,40)\Th  \putm(-12,40)\x \psline{->}(-12,35)(-12,25)}
\put(300,0){	\putlcf				\putc(0,-30){\fns\bf(MNP)}
	\putm(-12,20)\Th \psline{->}(12,35)(12,25)\putm(12,20)\Th 
	\putm(12,40)\Th \putm(-12,40)\Th \psline{->}(-12,35)(-12,25)}
\EP             \psset{linecolor=black}
\EC
\caption{Sample graphs for dressed coordinate transformations through
	second derivative order.}	\label{GEfig}	\VS-5
\end{figure}

For vector fields $\x$ the classical term is the Lie derivative, which amounts
to a change of coordinates. Its quantization yields an equivalence
transformation $D_\x$ of the form (\ref{EquiTrans}) so that the Kontsevich
product transforms covariantly under changes of coordinates only up to
gauge equivalence. For infinitesimal transformations
%% global construction becomes a non-trivial issue \cite{Kontsevich:1997vb}.
%%\BE	f\circ g = f * g - \frac{1}{24} \Th^{\mu\rho}\Th^{\nu\sigma}
%                   \partial_\rho \partial_\sigma 
%		\bigl( \ln \sqrt{\det(g-\cF)} \bigr)~[f_\mu * g_\nu]+\ldots
%%\EE
	\VS-5
\BE
	\6_t(f\kp_tg)
		=D_\x f\kp g+f\kp D_\x g-D_\x(f\kp g)\qquad
	\hbox{with}\qquad \dot x^\m=\x^\m.
\EE							\\[-12pt]
In fig. \ref{GEfig} we enumerate the graphs that contribute to infinitesimal
transformations (i.e. linear in $\x$) through second derivative order in
$\Th$. Note that the additional lines corresponding to derivatives acting
on $\x$ and $f$ can lead to different tensor structure, as indicated by
coefficients $A,\ldots,P$ for the terms with two derivatives on $\Th$'s,
plus an infinite number of additional Moyal-type contributions.
Through first derivative order
\BE
	D_\x=\x^\a\6_\a+\frac1{24}\x^\a_{\m\r}\Th^{\m\n}\6_\n
		\Th^{\r\b}\6_\a\6_\b+\cO(\6^2).
\EE 
At second derivative order the graphs define differential operators $D_\x$
containing (non-Moyal) terms with up to 
five derivatives, $D_\x f=\x^\a f_\a+\ldots
	+P \x^\a_{\r\s}\6_\n\Th^{\b\m}\6_\m\Th^{\r\g}
	\Th^{\d\n}\6_\n\Th^{\s\e}\,f_{\a\b\g\d\e}$ 
but many coefficients may be zero.

\del
Amounting to gauge operators
$D=1+D_1^\a\6_\a+D_2^{\a\b}
	\6_\a\6_\b+D_3^{\a\b\g}
	\6_\a\6_\b\6_\g+\ldots$
plus Moyal-type terms acting on $\x$ and $f$, like 
	$\d D_2^{\a\b}\!\!=a_1\,\Th^{\r\b}\x_{\n\r}^\m\6_\m\Th^{\n\a}$,
	~$\d D_3^{\a\b\g}\!\!=a_2\,\Th^{\r\b}\Th^{\s\g}\x_{\n\r\s}^\m\6_\m
	\Th^{\n\a}$, ~\ldots,
where
\BEA
	D_1^\a&=&\!\!   \x^\a+a\,\x_\n^\m\6_\m\Th^{\n\a}
		+A\,\x_\r^\m\6_\m\Th^{\b\n}\6_\n\Th^{\r\a}
		+C\,\x^\m_{\r\s}\Th^{\r\n}\6_\m\6_\n\Th^{\s\a}
		+E\,\x^\a_{\r\s}\6_\m\Th^{\r\n}\6_\n\Th^{\s\m}
\\	D_2^{\a\b}\!\!&=&\!\!	  b\,\x^\a_{\m\r}\Th^{\m\n}\6_\n\Th^{\r\b}
		+B\,\x_{\r\s}^\m\6_\m\Th^{\s\n}\6_\n\Th^{\r\a}
		+D\,\x^\m_{\r}\Th^{\a\n}\6_\m\6_\n\Th^{\r\b}
		+F\,\x^\a_{\r}\6_\m\Th^{\r\n}\6_\n\Th^{\m\b}
\nn\\&&\!\!\!\!	+H\,\x^\a_{\r\s\t}\Th^{\r\m}\Th^{\s\n}\6_\m\6_\n\Th^{\t\b}
		+K\,\x^\m_{\r\s\t}\6_\m\Th^{\r\a}\Th^{\s\n}\6_\n\Th^{\t\b}
\\	D_3^{\a\b\g}\!\!\!\!&=&\!\! c\,\x^\a_{\m}\Th^{\b\n}\6_\n\Th^{\m\g}
		+G\,\x^\a\6_\m\Th^{\b\n}\6_\n\Th^{\m\g}
		+I\,\x^\a_{\r\s}\Th^{\r\m}\Th^{\b\n}\6_\m\6_\n\Th^{\r\g}
		+L\,\x^\m_{\r\s}\6_\m\Th^{\r\a}\Th^{\b\n}\6_\n\Th^{\s\g}
\nn\\&&\!\!\!\!	+M\,\x^\a_{\r\s\t\k}\6_\n\Th^{\t\m}\6_\m\Th^{\r\b}
				\Th^{\k\n}\6_\n\Th^{\s\g}
\\	D_4^{\a\b\g\d}\!\!\!\!\!&=&\!\! 
		J\,\x^\a_{\r}\Th^{\b\m}\Th^{\g\n}\6_\m\6_\n\Th^{\r\d}
		+N\,\x^\a_{\r\s\t}\6_\n\Th^{\t\m}\6_\m\Th^{\r\b}
				\Th^{\g\n}\6_\n\Th^{\s\d}
\\	D_5^{\a\b\g\d\e}\!\!\!\!\!\!&=&\!\! 
		P\,\x^\a_{\r\s}\6_\n\Th^{\b\m}\6_\m\Th^{\r\g}
				\Th^{\d\n}\6_\n\Th^{\s\e}
	%	\hspace*{12cm}
\EEA

 \BI\item	coordinate transformation entails gauge transformation !
    \item	used by Cattaneo--Felder to ``globalize'' the $\kp$ product
 \EI
\enddel

\section{Open strings, Born--Infeld electrodynamics and non-commutativity}

In order to relate the %Let us discuss the origin of the 
Kontsevich product (\ref{KP}) to string theory we start with the
Polyakov action for closed strings moving
in a curved background with $2$-form field $B$. 
In conformal gauge 				\VS-5
\begin{equation}
  \label{Polyakov}
  S_P=\frac 1{2\pi\alpha'} \int_\S d^2z~ {\partial X^\mu \bar\partial X^\nu 
\Bigl(g_{\mu\nu}(X) + B_{\mu\nu}(X)\Bigr)},
\end{equation}					\\[-12pt]
where $X^\mu:\S \longrightarrow M$ maps the closed
world sheet $\S$ to the target manifold $M$. 
%%, which we take to be $\mathbb{R}^D$. 
Note that this action is invariant under the gauge transformation 
$\d_\L B=d\L$. % For what follows we set $2\pi\alpha'=1$.

When we consider open strings, we have to introduce world sheets with 
boundaries and specify a hypersurface in $M$, i.e. a D-brane, to which the
end points of  open strings are mapped. In the 
following we will only consider space-filling branes. 
By Stokes' theorem, (\ref{Polyakov}) is not gauge invariant
anymore, %that is 
$\int_\S  X^*\d_\L B=\int_{\6\S}X^*\L$, 
and we have to introduce a compensator field $A$ at the boundary, 
which turns out to be a $U(1)$ gauge field with field strength
$F=dA$. The associated action,			\VS-4
\begin{equation}
  \label{bdryaction}
   S_A=\int _{\6\S}X^* A=\int_{\6\S} dt~ \6_t X^\m A_\m(X)=\int_\S X^* F,
\end{equation}					\\[-12pt]
then restores gauge invariance of (\ref{Polyakov}) by setting
$\d_{\L,\l} A=-\frac1{2\p\a'}\L+d\l$. 
\del
The low-energy effective action derived from quantizing the system
(\ref{Polyakov}) plus (\ref{bdryaction}) encodes string amplitudes for
photon--graviton scattering such as:
\begin{center}			
\unitlength=2pt    \psset{unit=2pt} 
\BP(100,40)(-50,-20)	\ns
        \psccurve(-60,12)(-61.4,10)(-61.4,8)(-60,10)
        \psccurve(-60,-12)(-61.4,-10)(-61.4,-8)(-60,-10)
        \pscurve(-61.4,8)(-50,0)(-61.4,-8)
        \pscurve(-60,12)(-40,5)(-20,12) \pscurve(-60,-12)(-40,-5)(-20,-12)
        \pscurve(-18.6,9)(-30,0)(-18.6,-9)
        \psline(-18.6,9)(-20,12)  \psline(-18.6,-9)(-20,-12)
        \pscurve(-34,6)(-43,0)(-34,-6)                  \psline{->}(-6,0)(6,0)
                                \psline{->}(15,-10)(65,-10)
        \psdot(30,-10)\psdot(45,-10) \psdot(28,0)\psdot(49,4)
        \putc(-67,-11){$g_{\m\n}$}      \putc(28,4){$g_{\m\n}$}
        \putc(-67,11){$g_{\r\s}$}       \putc(49,8){$g_{\r\s}$}
        \putc(-15,-11){$A_{\a}$}       \putc(31,-15){$A_{\a}$}        
        \putc(-15,11){$A_{\b}$}        \putc(46,-15){$A_{\b}$}
        \pswedge[linestyle=dashed,linewidth=0.4pt](40,-10){25}{0}{180}
\EP
\end{center}
\enddel

As a consequence of  gauge symmetry % we expect that 
the effective action depends on the fields $A$ and $B$ only through the 
gauge invariant quantities $\cF=B+2\p\a'F$ and $H=dB=d\cF$.
% In fact, in the limit of constant curvature $\cF$ and metric $g$, 
For slowly varying fields $\cF$ and $g$ the 
effective theory on the D-brane is Born--Infeld electrodynamics
\cite{Fradkin:1985qd} governed by		\VS-4
\begin{equation}
  \label{BIaction}
  S_{BI} =\int_M d^Dx~ \sqrt{\det( \,g_{\m\n}+\cF_{\m\n}\,)}\ .
\end{equation}
				\\[-12pt]
% {\bf Remark:} Superstrings: $A~\to$ superpartner gaugino $\l$
%
Let us have a closer look at the quantization of (\ref{Polyakov}) and
(\ref{bdryaction}) on the upper half plane, conformally equivalent
to the disk. We split the embedding map into fluctuations around a
constant mode, $X^\m(z,\bar z) = x^\m + \z^\m(z,\bar z)$, and organize
the perturbative quantization in terms of a derivative expansion in
the background fields. Moreover, we regard the metric $g(x)$ and the
curvature $\cF(x)$ as a classical background in order to ensure
conformal invariance. 

The variation of the action requires the mixed Dirichlet--Neumann 
boundary condition						\VS-4
\BE%	g_{\m\n}(\6-\5\6)X^\n-\cF_{\m\n}(\6+\5\6)X^\n=
	g_{\m\n}\6_{t}X^\n
		-\cF_{\m\n}\6_{n}X^\n\bigr|_{\partial\S}=0\ ,
\EE							\\[-12pt]
% Here the derivatives are in tangent and normal direction to the world
% sheet boundary, respectively.
which leads to the following propagator for fluctuations at the
boundary ($\tau,\tau'\in  \partial\S$):				\VS-4
\begin{equation}
  \label{eq:fullprop}
  \bra \zeta^\mu(\tau)\,\zeta^\nu(\tau')\ket =
  - \frac{1}{2\pi} \Bigl\{G^{\mu\nu}(x) \ln |\tau-\tau'|^2 
  + i \pi \Theta^{\mu\nu}(x)~ \epsilon(\tau-\tau')\Bigr\},
\end{equation}							\\[-12pt]
where we introduced 
$G^{(\m\n)}+\Th^{[\m\n]}:=(g_{\m\n}+\cF_{\m\n})^{-1}$ and
the sign function $\epsilon(\tau)=\tau/|\tau|$. 
% $G^{\m\n}(x)$ is known as open
% string metric in view of its appearance in (\ref{eq:fullprop}) 
% \cite{Seiberg:1999vs}.

In the limit  when $g_{\m\n}$ vanishes (with $\cF_{\m\n}$ kept
finite) \cite{Seiberg:1999vs}, the action $S_P+S_A$ is topological. Only the
second part in (\ref{eq:fullprop}) survives, and the
non-commutative product on the D-brane world volume becomes
apparent. For constant backgrounds it is the
Moyal product. For varying backgrounds we notice, however, that the
Einstein equations for the background fields require $H=d\cF=0$ in the
topological limit \cite{Baulieu:2001fi}, i.e. $\cF$ is a symplectic form with
Poisson structure $\Theta = \cF^{-1}$. The resulting non-commutative
product is then the associative product (\ref{KP}) due to Kontsevich
\cite{Kontsevich:1997vb}.

\section{Associativity, cyclic invariance and effective actions}

{}From the string theory point of view, the assumption of $\Theta(x)$ being
a Poisson structure is not natural. The only condition on the background
fields should come from conformal invariance, or equivalently the classical
equations of motion. Therefore, it is preferable to define 
the non-commutative product without taking the
topological limit. It is clear that the first term in the propagator
(\ref{eq:fullprop}) should  play a secondary r\^ole in this
definition, which suggests to consider two (off-shell) vertex
operators at a distance $\tau'-\tau=1$ \cite{Herbst:2001ai}, i.e.\VS-12
\begin{equation}
  \label{defNCP}
  f(x)\circ g(x) :=
  \frac1{\sqrt{|g+\cF|}}~\int{\cal D}
  \z ~e^{-S[X=x+\z]}~f(X(0))~g(X(1)).
\end{equation}						\\[-7pt]
The Born--Infeld measure in the prefactor is cancelled by
(world sheet) $1$-loop diagrams. At higher derivative orders of the
background fields the measure gets corrected \cite{Okawa:1999cm}.  

% We expect that general open string scattering amplitudes can be
% expressed in terms of the product (\ref{defNCP}).

Let us comment on some properties of this product.	\VS-5
\BI
% \item Non-commutative product $\equiv$ \lila summing up leading \black 
% 	strong \lila$B/F$ \black field effects
\item  An immediate consequence of giving up
  on $\Theta(x)$ being Poisson is the \emph{loss of associativity}, so that a
  sum over different configurations of brackets will appear in open 
  string scattering amplitudes and in the effective action. In the
  topological limit, the non-commutative product (\ref{defNCP}) becomes 
  the Kontsevich product, up to gauge equivalence, 
  $D(f\, \kp\, g) = Df\circ Dg$, so that associativity is restored.\VS-5
\item  As was argued in \cite{cyclic}, the variational principle for the
  low-energy effective theory requires that the non-commutative
  product is \emph{cyclic}, i.e.
\begin{equation}
  \label{cyclic}
  \int_M \O ~f\circ g=\int_M \O ~f\cdot g\quad\hbox{and}\quad
  \int_M \O ~(f\circ g)\circ h=\int_M \O ~f\circ (g\circ h),
\end{equation}
  where $\Omega$ is a measure, which requires
$ %\begin{equation}
  \label{divergence}
  \partial_\m (\O~\Theta^{\m\n})=0.
$ %\end{equation}				\\[-12pt]
  From a string theory point of view the
  measure $\O$ is %not just any measure. It is 
the Born--Infeld
  measure that appeared in (\ref{BIaction}), i.e. $\O = \sqrt{|g+\cF|}$, %.
%  In fact \cite{AbouelsaoodCallan}, 
%%	Equation (\ref{divergence}) is the
and cyclicity follows from the  
generalized Maxwell equation associated with the Born--Infeld action
  (\ref{BIaction}): 
\BE	
	\6_\m(\sqrt{|g+\cF|}\Th^{\m\n})=0	\quad\iff\quad
	G^{\r\s}D_\r\cF_{\s\m}-\frac12\Th^{\r\s}H_{\r\s}{}^\l\cF_{\l\m}=0.
\EE
  This is in line with the assumption of a classical background, which
  ensures conformal invariance and, in
  particular, cyclic invariance of disk amplitudes.
  %and: associativity not sufficient; cyclicity (partly) fixes gauge 
  %	equivalence
  Notice that if we include the dilaton $\Ph$ in the background, the
  measure is modified to \,$e^{-\Ph}\sqrt{|g+\cF|}$.

  For Poisson structures the second condition in (\ref{cyclic})
  follows from associativity, and the first is due to 
	Connes--Flato--Sternheimer
  \cite{Connes:1992}. In fact, for any volume form $\O$
  subject to $\6_\m(\O\Th^{\m\n})=0$ %(\ref{divergence}) 
there exists a star-product that
  satisfies cyclic invariance (\ref{cyclic}) \cite{Felder:2000nc}. 
  However, in contrast to the physical context above, there is no
  canonical measure for Poisson structures.\VS-6
\item In \cite{Herbst:2001ai} an explicit computation of the product
  (\ref{defNCP}) was given to first derivative order, $\6\Th$, in the
  background field, but to all orders in $\Th$. In \cite{cyclic} it
  was shown that the cyclic invariance (\ref{cyclic}) uniquely fixes
  the non-commutative product to second derivative order, with the result\VS-4
\BE
	f\circ g=f\kp g-\frac1{24}
  \Th^{\m\r}\,\Th^{\n\s}\,\6_\r\6_\s(\log \O)
	\,f_\m\, g_\n .
\EE							\\[-12pt]
  The first contribution is the same expression (\ref{KP}) as the 
Kontsevich product but without the Poisson constraint on $\Th$ and the 
second is a gauge term that is needed to ensure cyclic invariance. 
\EI								\VS-3

If we want to use the non-commutative product (\ref{defNCP}) to
compute string S-matrix elements we have to impose on-shell conditions
not only on the background fields but also on the vertex operator
insertions. In the present context the vertex operators are functions,
$f(X)$, and thus the on-shell condition is the one for an open string
tachyon $T(x)$				\VS-4
\begin{eqnarray}
  \label{eq:eomtachyon}
  \square\, T = 
  \frac {1}{\O} ~\partial_\mu 
  \bigl(\O\, G^{\mu\nu}\partial_\nu T \bigr) = 
  -\frac{1}{\,\alpha'}\, T .
\end{eqnarray}				\\[-12pt]
This fixes the kinetic term for the low-energy effective
action. The
result is 	\VS-4
\BE
  S = - \frac 1{2g_o^2} \int_M \O~
      \Bigl\{ 
      G^{\mu\nu} \, \partial_\mu T \, \partial_\nu T -
      \frac {1}{\alpha'} ~T^2 - 
      {\sqrt \frac{8}{9\alpha'}} ~T \, (T \circ T) \Bigr\},
\EE				\\[-12pt]
where the cubic tachyon interaction was found 
%% \cut to first derivative order \endcut 
in \cite{Herbst:2002} by computing $3$-point amplitudes. 

%% \newpage

\section{Superstrings and non-anticommutative superspace}

The superstring in Green-Schwarz (GS) related formulations is an embedding
of a string in superspace. It thus appears natural that, in addition
to non-commutativity of space-time coordinates, there should be a
mechanism that deforms the anticommutation of the fermionic superspace
coordinates. Indeed such a mechanism exists. Independently of string
theory, special cases of non-anticommuting supercoordinates were
already considered by van Nieuwenhuizen and others in
\cite{Schwarz:1982pf} ($N=\frac{1}{2}$ SUSY,
see \cite{Seiberg:2003yz}). A more general ansatz was presented in
\cite{Klemm:2001yu}. 
After indications in \cite{Ferrara:2000mm,Kosinski:2000xu}
that similar structures originate from the superstring, this
could eventually be shown in \cite{Ooguri:2003qp} for a string in
four dimensions (with six dimensions compactified on a Calabi-Yau) and
was generalized in \cite{deBoer:2003dn} to ten dimensions. In both
cases a constant RR-field-strength was considered and turned out to be responsible
for the nonanticommutativity of the supercoordinates. 
% In addition, it was noticed that a gravitino background yields a 
% nonvanishing commutator between bosonic and fermionic coordinates.
The calculations where performed in different versions of the covariant 
superstring 
\cite{Berkovits:2002zk,Grassi:2003cm,Berkovits:1994wr}. 
This non-(anti)commutativity can again be implemented via a star product,
now on superspace (see \cite{Tyutin:2001iz} and references therein).
For non-constant background fields (but in the topological limit),
this corresponds to a graded generalization of Kontsevich's associative
star product. A derivation from a $\sigma$-model with super-targetspace
along the lines of Cattaneo and Felder \cite{Cattaneo:2000fm} was
presented in \cite{Chepelev:2003ga}. The effect of a constant RR-potential (not field strength) on the deformation of the bosonic space was already studied in \cite{Cornalba:2002cu}. %% newer things?
In the following we sketch how non-anticommutativity of superspace 
%coordinates 
arises from the Berkovits pure spinor superstring %, as it was presented in 
\cite{deBoer:2003dn}. 
% To that end, let us quickly introduce the pure spinor formalism, 
% starting from the GS string.

%\paragraph{From GS to pure spinor formalism} 
Although we will consider an open string with type I supersymmetry, we want 
to couple it to the type II bulk fields (see e.g. \cite{Cornalba:2002cu}). In particular the RR-fields belong to the bulk and will take over the role of the B-field in the fermionic case.
It is therefore necessary to embed the string into a type II superspace with 
coordinates $x^{M}=(x^{m},\theta^{\mu},\hat{\theta}^{\hat{\mu}})$.
% Note that, for ``historic'' reasons, we change the notation for the 
% indices. Bosonic indices will no longer be denoted by Greek letters. 
In this section Greek letters will be reserved for fermionic indices
while bosonic indices are denoted by Latin letters.
In conformal gauge, the GS action in flat background reads	\VS-3
\begin{eqnarray}			S_{GS} & \equiv & \textstyle
	\int \rm{d}^2z\quad\frac{1}{2}\Pi_{z}^{a}\eta_{ab}\Pi_{\bar{z}}^{b}
	+\mc{L}_{WZ}\qquad\textrm{(conformal gauge)}\nonumber\\
\mc{L}_{WZ} & \equiv & \textstyle
	-\frac{1}{2}\Pi_{z}^{a}\left(\theta\gamma_{a}\bar{\partial}\theta
	-\hat{\theta}\gamma_{a}\bar{\partial}\hat{\theta}\right)+\frac{1}{2}
	(\theta\gamma^{a}\partial\theta)(\hat{\theta}
	\gamma_{a}\bar{\partial}\hat{\theta})-(z\leftrightarrow\bar{z}),
\end{eqnarray}	
where $\Pi_{z/\bar z}^a$ are the supersymmetric momenta. They can be 
described as the pullback of the bosonic part of the 
supervielbein\vspace{-.6cm} 
\BE	\mathbb{E}^{A}\equiv\textrm{d}x^{M}\mathbb{E}_{M}\hoch{A}
	\stackrel{\rm{flat}}{=}\big(\overbrace{
	\textrm{d}x^{a}+\textrm{d}\theta\gamma^{a}\theta+\textrm{d}
	\hat{\theta}
	\gamma^{a}\hat{\theta}}^{\Pi^{a}}\;,\;\textrm{d}
	\theta^{\alpha}\;,\;\textrm{d}\hat{\theta}^{\hat{\alpha}}\big)
\EE 
to the worldsheet. Letters from the beginning of the alphabet shall denote 
``flat indices'' (with respect to the local frame), while letters from the 
end of the alphabet will denote ``curved indices''. This distinction is more 
relevant for the curved background to be discussed later. As usual, 
$\gamma^a_{\alpha\beta}$ denotes the off-diagonal chiral block of the 
10-dimensional Dirac gamma matrix $\Gamma^a$, in a representation where it 
is real and symmetric (i.e. graded antisymmetric) in the indices $\alpha$ 
and $\beta$.

The Wess-Zumino term $\mc{L}_{WZ}$ is responsible for the existence of a 
local fermionic symmetry, the $\kappa$-symmetry. Indeed, the theory contains 
a number of fermionic constraints $d_{z\a}$, $\hat d_{\bar z \a}$. Only half 
of each set, however, is first class and the constraint algebra is therefore 
not closed:
\begin{eqnarray}
	\left\{ d_{z\alpha}(\sigma),d_{z\beta}(\sigma')\right\}  & \propto 
	& 2\gamma_{\alpha\beta}^{a}\Pi_{za}\delta(\sigma-\sigma').
				\label{constraint-algebra} 
\end{eqnarray}
Being a spinor in an irreducible representation, $d_{z\a}$ cannot covariantly 
be separated into first and second class and thus does not allow covariant 
quantization. A long struggle to overcome this problem resulted in the 
invention of the pure spinor string \cite{Berkovits:2000fe, Berkovits:2002zk} 
as an alternative formalism. 

Berkovits' pure spinor formalism has two basic ingredients. The first is a 
free action of the form					\VS-3
\begin{eqnarray}	S_{free} & = & \textstyle
	\int{\rm d}^2z\quad\frac{1}{2}\partial x^{m}\eta_{mn}\bar{\6}x^{n}
	+\bar{\partial}\theta^{\alpha}p_{z\alpha}
	+\6\hat{\theta}^{\hat{\alpha}}\hat{p}_{\bar{z}\hat{\alpha}}\nonumber
\\	 & = & \textstyle
	\int{\rm d}^2z\quad\frac{1}{2}\Pi_{z}^{a}\eta_{ab}\Pi_{\bar{z}}^{b}
	+\mc{L}_{WZ}+\bar{\partial}\theta^{\alpha}d_{z\alpha}
	+\partial\hat{\theta}^{\hat{\alpha}}\hat{d}_{\bar{z}\hat{\alpha}}
	\label{Sfree} ,
\end{eqnarray}
where $p_{z\alpha}$, $\hat p_{\bar z\a}$ are independent variables and 
$
	d_{z\alpha} \equiv p_{z\alpha}-(\gamma_{a}\theta)_{\alpha}
	\bigl(\partial x^{a}
	-\tfrac{1}{2}\theta\gamma^{a}\partial\theta
	-\tfrac{1}{2}\hat{\theta}\gamma^{a}\partial\hat{\theta}\bigr)
$
and its hatted counterpart have the same algebra as the constraints of the 
GS-string. In addition, this action
 coincides classically with the GS-action for $d_{z\alpha}
	=\hat{d}_{\bar z\hat{\alpha}}=0$. The second basic ingredient are 
the BRST operators   		\VS-8
\BE\textstyle
	Q=\oint\textrm{d}z~\lambda^{\alpha}d_{z\alpha},
	\quad\hat{Q}=\oint\textrm{d}\bar{z}
	~\hat{\lambda}^{\hat{\alpha}}\hat{d}_{\bar{z}\hat{\alpha}},
\EE		\\[-18pt]
which implement in some sense the constraints $d_{z\alpha}=\hat{d}_{\bar 
z\hat{\alpha}}=0$ cohomologically. $\lambda^\a$ and $\hat\lambda^{\hat\a}$ 
are ghost fields of even parity. Containing also second class constraints, 
the above BRST operators fail to be nilpotent in general. This can be 
repaired by constraining the ghost fields to be so-called pure spinors, 
obeying
$
	\lambda\gamma^a\lambda=\hat\lambda\gamma^a\hat\lambda=0\,.
$
Like the fermionic coordinates, the ghost fields should be left and 
right-moving respectively and one thus adds the corresponding ghost term to 
the free action (\ref{Sfree}):
$%\begin{eqnarray}			\textstyle
	S_{ps}  =  S_{\rm free}+\int d^2z
	\quad\bar{\partial}\lambda^{\alpha}\omega_{z\alpha}+\partial
	\hat{\lambda}^{\hat{\alpha}}\omega_{\bar z\hat{\alpha}}+L_{z\bar{z}a}
	(\lambda\gamma^{a}\lambda)+\hat{L}_{z\bar{z}a}(\hat{\lambda}
	\gamma^{a}\hat{\lambda}).
$ %\end{eqnarray}\ns\rm
The implementation of the pure spinor constraints with the help of Lagrange 
multipliers immediately reveals (by varying with respect to the ghost) a 
gauge symmetry of the antighosts of the form 
$\delta_{(\mu)}\omega_{z\alpha}=\mu_{za}(\gamma^{a}\lambda)_{\alpha}$ which 
corresponds to the (first-class) pure-spinor constraints. Because the field 
equations are basically free, one gets free field operator products after 
quantization. For the antighost field, this statement is restricted to gauge 
invariant operators like the ghost current or the Lorentz current. Apart 
from the central charges, their OPEs look as if there was no pure spinor 
constraint. To determine the central charges, one has to solve the 
constraint once (see e.g. \cite{Berkovits:2002zk}).

In order to complete the description for the open string, we still need 
boundary conditions. For vanishing background a natural choice is to set 
$\theta=\hat\theta$ and $\lambda=\hat\lambda$ at the boundary. This can 
be implemented by the variation of a boundary term that one should add 
to the action. The precise form of this boundary term is fixed by 
%requiring 
$N=1$ supersymmetry, BRST invariance and the antighost gauge symmetry. As the 
final form of the boundary action is quite lengthy and not very illuminating, 
we refer to \cite{Berkovits:2002ag} for further details.

%\paragraph{Pure spinor string in general background}
% originally: As mentioned in the beginning, the object of interest is an open string in 
% a general background, containing boundary and bulk fields. 
% The string action will thus consist of a bulk part 
% which just has the same form as a closed string in general background and in addition
% a boundary part which is only present for the open string.\enlargethispage*{1cm} 
% Max: As in the bosonic case, the string action will consist of a bulk part
% and additional boundary terms.
The open string in a general background of bulk and boundary fields consists of a bulk part of the same form as a closed string in general background and an additional boundary part. 
The closed pure spinor superstring in general background was studied first
 by Berkovits and Howe in \cite{Berkovits:2001ue}. Already at classical 
level, conservation and nilpotency of the BRST charges implement the 
type II supergravity constraints. Those, in turn, guarantee 1-loop quantum 
conformal invariance of the theory \cite{Bedoya:2006ic}. The presentation of 
the bulk part in the following is based on \cite{Guttenberg:2007th}. The 
starting point is the most general classically conformally invariant 
action:					\VS-4
\begin{eqnarray}\txt
	\hspace{-1cm} S_{bulk} & = & 	
	\int d^{2}z\quad\frac{1}{2}
	\partial x^{M}(\mathbb{G}_{MN}(\xfull)+\mathbb{B}_{MN}(\xfull))
	\bar{\partial}x^{N}+\bar{\partial}x^{M}\mathbb{E}_{M}
	\hoch{\bs{\alpha}}(\xfull)\:\dP_{z\bs{\alpha}}+\partial x^{M}
	\mathbb{E}_{M}\hoch{\hat{\bs{\alpha}}}(\xfull)\:\hat{\dP}_{\bar{z}
	\hat{\bs{\alpha}}}+\nonumber \txt\\
 &  & +\dP_{z\bs{\alpha}}\RR^{\bs{\alpha}\hat{\bs{\beta}}}(\xfull)
	\:\hat{\dP}_{\bar{z}\hat{\bs{\beta}}}+\ce^{\bs{\alpha}}
	\mathbb{C}_{\bs{\alpha}}\hoch{\bs{\beta}\hat{\bs{\gamma}}}(\xfull)
	\:\be_{z\bs{\beta}}\hat{\dP}_{\bar{z}\hat{\bs{\gamma}}}
	+\hat{\ce}^{\hat{\bs{\alpha}}}\hat{\mathbb{C}}_{\hat{\bs{\alpha}}}
	\hoch{\hat{\bs{\beta}}\bs{\gamma}}(\xfull)\:\hat{\be}_{\bar{z}
	\hat{\bs{\beta}}}\dP_{z\bs{\gamma}}+\nonumber \\
 &  & +\Bigl(\bar{\partial}\ce^{\bs{\beta}}+\ce^{\bs{\alpha}}\bar{\partial}
	x^{M}\Omega_{M\bs{\alpha}}\hoch{\bs{\beta}}(\xfull)\Bigr)
	+\Bigl(\partial\hat{\ce}^{\hat{\bs{\beta}}}+\hat{\ce}^{\hat{\bs{\a}}}
	\partial x^{M}\hat{\Omega}_{M\hat{\bs{\alpha}}}\hoch{\hat{\bs{\beta}}}
	(\xfull)\Bigr)\hat{\be}_{\bar{z}\hat{\bs{\beta}}}+\nonumber \\
 &  & +\frac{1}{2}L_{z\bar{z}a}(\ce\gamma^{a}\ce)+\frac{1}{2}\hat{L}_{\bar{z}z
	\hat{a}}(\hat{\ce}\gamma^{\hat{a}}\hat{\ce})+\ce^{\bs{\alpha}}
	\hat{\ce}^{\hat{\bs{\alpha}}}\mathbb{S}_{\bs{\alpha}\hat{\bs{\alpha}}}
	\hoch{\bs{\beta}\hat{\bs{\beta}}}(\xfull)\:\be_{z\bs{\beta}}
	\hat{\be}_{\bar{z}\hat{\bs{\beta}}}\label{eq:BiBaction}
\end{eqnarray}			\\[-12pt]
The variable $\xfull$ contains $x^{m},\theta^{\bs{\mu}}$ and 
$\hat{\theta}^{\hat{\bs{\mu}}}$. In addition to the action, we need the two 
BRST operators. In principle they could contain background fields as well, but 
it is always possible to reparametrize $d_{z\a}$ and $\hat d_{\bar z\hat\a}$ 
such that they have the same form as in the flat case.
\del
\BE\textstyle
	Q=\oint\textrm{d}z\quad\ce^{\bs\alpha}d_{z\bs\alpha},
	\qquad\hat{Q}=\oint\textrm{d}\bar{z}
	\quad\hat{\ce}^{\hat{\bs\alpha}}\hat{d}_{\bar{z}\hat{\bs\alpha}}\quad .
\EE
\enddel
Consistency of the equations of motion with the pure spinor constraints 
requires that the background fields $\Omega_{M\bs{\alpha}}\hoch{\bs\beta}$ and 
$\hat\Omega_{M\bs{\hat\alpha}}\hoch{\bs{\hat\beta}}$ are each a sum of a 
spinorial Lorentz-transformation and dilatation in the last two indices. They 
can thus be regarded as Lorentz plus scale connections. This property also 
establishes the antighost gauge symmetry in the general case. BRST invariance 
of the action requires that the symmetric two-tensor is of the form 
$\mathbb G_{MN}=\mathbb E_M\hoch{a}\eta_{ab}\mathbb E_N\hoch{b}$.
The background fields $\mathbb E_M\hoch{a}$, $\mathbb E_M\hoch{\bs\a}$ and 
$\mathbb E_M\hoch{\hat{\bs\a}}$ can then be combined to a single object 
$\mathbb E_M\hoch{A}$ and regarded as supervielbein. BRST invariance and 
nilpotency of the BRST transformations put several restrictions on the background 
fields which turn out to be equivalent to the type II supergravity constraints 
\cite{Berkovits:2001ue,Bedoya:2006ic,Guttenberg:2007th}. 
% All those constraints are still necessary 
% (but not any longer sufficient) in the case of open strings. 

For the moment, we restrict ourselves to a glance at the propagator. 
I.e., we are interested in the quadratic part of the action and do not yet 
need all the constraints. 
Expanding the coordinates around a constant zero mode, restricting to 
vanishing zero mode for the fermionic coordinates and the ghosts, choosing a 
parametrization which corresponds to the  WZ-gauge and restricting 
to the quadratic part, one arrives at		\VS-7
\begin{eqnarray}
\hspace{-1cm}\bei{S_{qu}}{\q\theta=\q\ce=0}
	= &  & \txt\hspace{-.5cm}\int d^{2}z~\frac{1}{2}
	\partial\zeta^{m}\left(e_{m}\hoch{a}(\q{\xboson})\eta_{ab}e_{n}
	\hoch{b}(\q{\xboson})+B_{mn}(\q{\xboson})\right)\bar{\partial}
	\zeta^{n}+\dP_{z\bs{\alpha}}\rr^{\bs{\alpha}\hat{\bs{\beta}}}
	(\q{\xboson})\:\hat{\dP}_{\bar{z}\hat{\bs{\beta}}}
 \\
&&	\HS-63
	+\bar{\partial}\zeta^{m}\psi_{m}\hoch{\bs{\alpha}}(\q{\xboson})
	\:\dP_{z\bs{\alpha}}+\bar{\partial}\zeta^{\bs{\mu}}
	\delta_{\bs{\mu}}\hoch{\bs{\alpha}}\dP_{z\bs{\alpha}}
	+\partial\zeta^{m}\hat{\psi}_{m}\hoch{\hat{\bs{\alpha}}}(\q{\xboson})
	\:\hat{\dP}_{\bar{z}\hat{\bs{\alpha}}}
	+\partial\zeta^{\hat{\bs{\mu}}}\delta_{\hat{\bs{\mu}}}
	\hoch{\hat{\bs{\alpha}}}\hat{\dP}_{\bar{z}\hat{\bs{\alpha}}}
	+\nonumber 
%\\&  & \qquad+
	\bar{\partial}\ce^{\bs{\beta}}\be_{z\bs{\beta}}+\partial
	\hat{\ce}^{\hat{\bs{\beta}}}\hat{\be}_{\bar{z}\hat{\bs{\beta}}}
\end{eqnarray}							\\[-14pt]
with $x^{M}(z,\bar{z}) = \q{x}^{M}+\zeta^{M}(z,\bar{z})$.
At this stage it becomes visible that the Ramond-Ramond (RR) fields 
$\rr^{\bs\a\hat{\bs\b}}$ will enter the propagator between the fermionic 
coordinates.
% , while the gravitinos $\psi_m\hoch{\bs\a}$ and 
% $\hat\psi_m\hoch{\hat{\bs\a}}$ will enter the propagator between bosonic 
% and fermionic coordinates. 
This observation was made for constant RR-fields in \cite{Ooguri:2003qp} 
for four dimensions (with six compactified on a Calabi-Yau) and in 
\cite{deBoer:2003dn} for ten dimensions. The associated 
anticommutation relations were found to be			\VS-6
\BE
	% [x^m,x^n]\propto\Theta^{mn},\quad 
	\{\theta^{\bs\a},\hat
	\theta^{\hat{\bs\b}}\}\propto\rr^{\bs\a\hat{\bs\b}}
	\label{nonanticommutativity}.
\EE						\\[-18pt]
Turning on the field strength $\cF$ modifies the boundary conditions for all 
world sheet fields and also leads to a RR background
dependent shift in the noncommutativity parameter $\Theta^{mn}$
\cite{Cornalba:2002cu}.

For general backgrounds, one needs to check the consistency of the boundary 
action with the bulk BRST transformations and the pure spinor constraints. 
Already for the open pure spinor string in an open string background this 
is a long story, which was discussed by Berkovits and Pershin in 
\cite{Berkovits:2002ag}. In addition to the boundary term that was mentioned 
before they add the integrated open string vertex operator of the form\VS-7
\BE
	V\propto\int d\tau\quad \dot\theta_+^{\bs\a}A_{\bs\a}(x,\theta_+)
	+\Pi_+^mB_m(x,\theta_+)+d_{\bs\a}^+W^{\bs\a}(x,\theta_+)+
	\frac12(N_+)_{\bs\a}^{\bs\b}(\gamma F)^{\bs\a}_{\bs\b}(x,\theta_+)
\EE							\\[-15pt]
to the action. The worldline fields with index '+' are just suitable linear 
combinations of the left and rightmovers and 
$(N_+)_{\bs\alpha}^{\bs\b}\propto \ce_+^{\bs\b}\be^+_{\bs\a}$. The objects 
$A_{\bs\a}$, $B_m$, $W^{\bs\a}$ and $F^{\bs\a\bs\b}$ are $N=1$ background 
superfields. The consistency requirements of the boundary action with BRST 
invariance and the pure spinor constraint leads to the field equations of 
supersymmetric Born--Infeld for these background superfields. 

In order to generalize the result (\ref{nonanticommutativity}) to non-constant
bulk fields one has to become yet more general, combining the boundary part 
$V$ with the bulk action (\ref{eq:BiBaction}) and 
studying the consistent boundary 
conditions and field equations. This is work in progress.

%%%
% \item	Most tedious part: solution of constraints, $\th$-expansion
% \\	(cf. Dimitrios' talk)

\section{Conclusion}

In this note we gave an introduction to the Kontsevich product and discussed
our proposal for a generalization to the non-associative case. We established
cyclicity through second derivative order, which allows for the 
non-commutative product to be used in the construction of effective actions.
We checked that our previous results \cite{cyclic} generalize to
non-constant dilaton backgrounds, with the only modification being the
prefactor $\exp(-\Ph)$ in the measure. We also reviewed the existing
results and ideas about generalizations to superstrings, which have been
investigated so far for constant background fields.

There is a number of obvious directions for further work. For the bosonic
string a non-commutative generalization of the gauge field effective action
should be constructed, which presumably is related to derivative corrections
to the measure. The non-abelian case should also have interesting implications
for commutative non-abelien Born-Infeld actions.
A quite demanding task will be the generalization of our results to 
superstrings in curved $RR$ and $B$-field backgrounds.
On the more mathematical side, it would be interesting to establish cyclicity
to all orders in the derivative expansion and if possible explicitly
construct the non-associative product.

\del
\BI
\item	extend to higher order of give proof to all orderns: 
	* is there a non-topological version formality?
\item	Photon / non-abelien non-commutative effective action: 
\\	* is there a generalization of non-commutative gauge 
		invariance and the Seiberg-Witten map?
\\	* guess/compute derivative corrections to (BI) measure? 
		cf. \cite{Okawa:1999cm}
\\	* general coordinate covariance (with $f\to\cD f$
	discussed above) should give further constraints;
\\	(internal remark: suggestion by John Madore in discussion of my talk:
	use Vielbein basis in BI expansion) 
\item	study deformation of superspace by (non-constant) RR (and bosonic 
	backgrounds $\cF,\Ph$) ...
\item	apply to whatever may come to your mind
\EI
\enddel

\begin{acknowledgement}
We would like to thank Giovanni Felder, Karl-Georg Schlesinger and Ricardo 
Schiappa for helpful discussions. S.G. was supported by the European Network 
on Random Geometry, EEC Grant No. MRTN-CT-2004-005616. M. K. acknowledges 
support by the Austrian Research Funds FWF grant P19051-N16.
\end{acknowledgement}

%\goodbreak


\begin{thebibliography}{[1]}		\def\T#1{{\it #1}}


%\cite{Kontsevich:1997vb}	1
\bibitem{Kontsevich:1997vb}%
M.~Kontsevich,
\T{Deformation quantization of Poisson manifolds, I,}
~[arxiv:q-alg/9709040].
%%CITATION = Q-ALG 9709040;%%

%\cite{Sternheimer:1998yg}	2
\bibitem{Sternheimer:1998yg}%
D.~Sternheimer,
\T{Deformation quantization: Twenty years after,}	%\cut
AIP Conf.\ Proc.\  {\bf 453} (1998) 107			\\{}%\endcut
[arXiv:math.qa/9809056].			
%%CITATION = MATH-QA 9809056;%%

%\cite{Schomerus:1999ug}	3
\bibitem{Schomerus:1999ug}%
V.~Schomerus,
\T{D-branes and deformation quantization,}
JHEP {\bf 9906} (1999) 030
~[arxiv:hep-th/9903205].
%%CITATION = HEP-TH 9903205;%%

%\cite{Seiberg:1999vs}		4
\bibitem{Seiberg:1999vs}%
N.~Seiberg, E.~Witten,
\T{String theory and noncommutative geometry,}
JHEP {\bf 9909} (1999) 032
~[hep-th/9908142].
%%CITATION = HEP-TH 9908142;%%



%\cite{Cattaneo:2000fm}		5
\bibitem{Cattaneo:2000fm}%
A.~S.~Cattaneo, G.~Felder,
\T{A path integral approach to the Kontsevich quantization formula,}
Commun.\ Math.\ Phys.\  {\bf 212} (2000) 591
~[arxiv:math.qa/9902090].
%%CITATION = MATH.QA 9902090;%%

\bibitem{Cornalba:2001sm}
  L.~Cornalba and R.~Schiappa,
  {\it Nonassociative star product deformations for D-brane worldvolumes in
  curved backgrounds,}
  Commun.\ Math.\ Phys.\  {\bf 225} (2002) 33
  [arXiv:hep-th/0101219].


%\cite{Herbst:2001ai}		6
\bibitem{Herbst:2001ai}%
M.~Herbst, A.~Kling, M.~Kreuzer,
\T{Star products from open strings in curved backgrounds,}
JHEP {\bf 0109} (2001) 014
~[arXiv:hep-th/0106159].

%%CITATION = HEP-TH 0106159;%%
%\cite{Schoikhet:1999}		7
\bibitem{Schoikhet:1999}%
B.~Shoikhet,
\T{On the cyclic formality conjecture,}
~[arXiv:math.qa/9903183].

%\cite{Felder:2000nc}		8
\bibitem{Felder:2000nc}%
G.~Felder and B.~Shoikhet,
\T{Deformation quantization with traces,}
~[arXiv:math.qa/0002057].
%%CITATION = MATH-QA 0002057;%%

							%	9
\bibitem{cyclic}M.~Herbst, A.~Kling, M.~Kreuzer,
	\T{	Cyclicity of non-associative products on D-branes,}
   J. High Energy Physics  \bf0403 \rm(2004) 003 ~[arxiv:hep-th/0312043]

\bibitem{DeWiLe83}De Wilde, M. and Lecomte, P.B.A. 	%	10
	{\it Existene of
	star-products and of formal deformations of the Poisson Lie algebra of
	arbitrary symplectic manifolds,} Lett. Math. Phys. 7 (1983) 487-496

\bibitem{Fedosov}Fedosov B.B. {\it Formal quantization} in %	11
	{\it Some topics 
	of modern mathematics and their applications to problems of 
	mathematical physics} (Moscow, 1985) 129;
	{\it Quantization and index}, Dokl. Akad. Nauk SSSR {\bf 291} (1986) 
	82


%%%% next 3 entries added by Manfred at November 29, 2007

%\cite{Fradkin:1985qd}			%	12
\bibitem{Fradkin:1985qd}%
  E.~S.~Fradkin, A.~A.~Tseytlin,
  \T{Nonlinear electrodynamics from quantized strings,}
  Phys.\ Lett.\  B {\bf 163} (1985) 123.
  %%CITATION = PHLTA,B163,123;%%

%\cite{Baulieu:2001fi}			%	13
\bibitem{Baulieu:2001fi}%
  L.~Baulieu, A.~S.~Losev and N.~A.~Nekrasov,
  \T{Target space symmetries in topological theories. I,}
  JHEP {\bf 0202} (2002) 021
  ~[arXiv:hep-th/0106042].
  %%CITATION = JHEPA,0202,021;%%

%\cite{Okawa:1999cm}			%	14
\bibitem{Okawa:1999cm}%
Y.~Okawa,
\T{Derivative corrections to Dirac--Born--Infeld Lagrangian and 
	non-commutative gauge theory,}
Nucl.\ Phys.\ B {\bf 566} (2000) 348
[arXiv:hep-th/9909132].
%%CITATION = HEP-TH 9909132;%%

\del
\bibitem{AbouelsaoodCallan}%			15
  A.~Abouelsaood, C.~G.~.~Callan, C.~R.~Nappi, S.~A.~Yost,
  \T{Open Strings In Background Gauge Fields,}
  Nucl.\ Phys.\  B {\bf 280} (1987) 599;
  C.~G.~.~Callan, C.~Lovelace, C.~R.~Nappi, S.~A.~Yost,
  \T{String Loop Corrections To Beta Functions,}
  Nucl.\ Phys.\  B {\bf 288} (1987) 525.
\enddel

%\cite{Connes:1992}				%	16
\bibitem{Connes:1992}%
A.Connes, M.Flato, D.Sternheimer,
\T{Closed star-products and cyclic cohomology,}
Lett. Math. Phys. {\bf 24} (1992) 1

\bibitem{Herbst:2002}%	17
M.~Herbst, A.~Kling and M.~Kreuzer,
\T{Non-commutative tachyon action and D-brane geometry,}
JHEP {\bf 0208} (2002) 010
~[arXiv:hep-th/01mmnnn].

\del

%\cite{Ardalan:1999ce}				%	
\bibitem{Ardalan:1999ce}%
F.~Ardalan, H.~Arfaei and M.~M.~Sheikh-Jabbari,
\T{Noncommutative geometry from strings and branes,}
JHEP {\bf 9902} (1999) 016
~[arxiv:hep-th/9810072].
%%CITATION = HEP-TH 9810072;%%

%\cite{Schaller:1994es}				
\bibitem{Schaller:1994es}
P.~Schaller and T.~Strobl,
\T{Poisson structure induced (topological) field theories,}
Mod.\ Phys.\ Lett.\ A {\bf 9} (1994) 3129
~[arXiv:hep-th/9405110];~
%%CITATION = HEP-TH 9405110;%%
%\cite{Schaller:1995xk}
%\bibitem{Schaller:1995xk}
P.~Schaller and T.~Strobl,
\T{Introduction to Poisson sigma-models,}
~[arXiv:hep-th/9507020.
%%CITATION = HEP-TH 9507020;%%

%\cite{Cornalba:2002sm}
\bibitem{Cornalba:2002sm}
L.~Cornalba and R.~Schiappa,
\T{Nonassociative star product deformations for D-brane worldvolumes 
	in curved backgrounds,}
Commun.\ Math.\ Phys.\  {\bf 225} (2002) 33
~[arXiv:hep-th/0101219].
%%CITATION = HEP-TH 0101219;%%

%\cite{Andreev:2001xx}
\bibitem{Andreev:2001xx}
O.~Andreev,
\T{More about partition function of open bosonic string in background 
fields and string theory effective action,}
Phys.\ Lett.\ B {\bf 513} (2001) 207
~[arXiv:hep-th/0104061].
%%CITATION = HEP-TH 0104061;%%

%\cite{Wyllard:2000qe}
\bibitem{Wyllard:2000qe}
N.~Wyllard,
\T{Derivative corrections to D-brane actions with constant background fields,}
Nucl.\ Phys.\ B {\bf 598} (2001) 247
~[arXiv:hep-th/0008125].
%%CITATION = HEP-TH 0008125;%%

%\cite{Fotopoulos:2001pt}
\bibitem{Fotopoulos:2001pt}
A.~Fotopoulos,
\T{On (alpha')**2 corrections to the D-brane action for non-geodesic 
	world-volume embeddings,}
JHEP {\bf 0109} (2001) 005
~[arXiv:hep-th/0104146].
%%CITATION = HEP-TH 0104146;%%

%\cite{Das:2001xy}
\bibitem{Das:2001xy}
S.~R.~Das, S.~Mukhi and N.~V.~Suryanarayana,
\T{Derivative corrections from noncommutativity,}
JHEP {\bf 0108} (2001) 039
~[arXiv:hep-th/0106024].
%%CITATION = HEP-TH 0106024;%%


%\cite{Wyllard:2001ye}
\bibitem{Wyllard:2001ye}
N.~Wyllard,
\T{Derivative corrections to the D-brane Born--Infeld action: 
	Non-geodesic embeddings and the Seiberg-Witten map,}
JHEP {\bf 0108} (2001) 027
[arXiv:hep-th/0107185].
%%CITATION = HEP-TH 0107185;%%

%\cite{Pal:2001xp}
\bibitem{Pal:2001xp}
S.~S.~Pal,
\T{Derivative corrections to Dirac--Born--Infeld and Chern--Simon actions 
	from non-commutativity,}
Int.\ J.\ Mod.\ Phys.\ A {\bf 17} (2002) 1253
[arXiv:hep-th/0108104].
%%CITATION = HEP-TH 0108104;%%

\bibitem{Penkava:1998xx}
M.~Penkava and P.~Vanhaecke,
\T{Deformation quantization of polynomial Poisson algebras,}
[arXiv:math.QA/9804022].
%%CITATION = MATH.QA/9804022;%%

%\cite{Zotov:2001ec}
\bibitem{Zotov:2001ec}
A.~Zotov, \T{On relation between Moyal and Kontsevich quantum products. 
	Direct evaluation up to the $\hbar^3$-order,}
Mod.\ Phys.\ Lett.\ A {\bf 16} (2001) 615
~[arXiv:hep-th/0007072].
%%CITATION = HEP-TH 0007072;%%

%\cite{Dito:2002dr}
\bibitem{Dito:2002dr}
G.~Dito and D.~Sternheimer,
\T{Deformation auantization: Genesis, developments and metamorphoses,}
~[arXiv:math.qa/0201168].
%%CITATION = MATH-QA 0201168;%%

\bibitem{Gaberdiel:1997ia}
M.~R.~Gaberdiel and B.~Zwiebach,
\T{Tensor constructions of open string theories I: Foundations,}
Nucl.\ Phys.\ B {\bf 505} (1997) 569
~[arXiv:hep-th/9705038].

%\cite{Zwiebach:1992ie}
\bibitem{Zwiebach:1992ie}
B.~Zwiebach,
\T{Closed string field theory: Quantum action and the B-V master equation,}
Nucl.\ Phys.\ B {\bf 390} (1993) 33
~[arXiv:hep-th/9206084].
%%CITATION = HEP-TH 9206084;%%

%\cite{Alexandrov:1995kv}
\bibitem{Alexandrov:1995kv}%
M.~Alexandrov, M.~Kontsevich, A.~Schwartz and O.~Zaboronsky,
\T{The Geometry of the master equation and topological quantum field theory,}
Int.\ J.\ Mod.\ Phys.\ A {\bf 12} (1997) 1405
~[arXiv:hep-th/9502010].
%%CITATION = HEP-TH 9502010;%%

\bibitem{Kajiura:2003ax}
H.~Kajiura,
\T{Noncommutative homotopy algebras associated with open strings,}
arXiv:math.qa/0306332.


\enddel

\bibitem{Schwarz:1982pf}%			%	18
  J.~H.~Schwarz and P.~Van Nieuwenhuizen,
  \T{Speculations Concerning A Fermionic Substructure Of Space-Time,}
  Lett.\ Nuovo Cim.\  {\bf 34} (1982) 21.
  %%CITATION = NCLTA,34,21;%%

%\cite{Seiberg:2003yz}			% 19
\bibitem{Seiberg:2003yz}%
N.~Seiberg,
\T{Noncommutative superspace, N = 1/2 supersymmetry, field theory and 
	string theory,}
JHEP {\bf 0306} (2003) 010
~[arXiv:hep-th/0305248].
%%CITATION = HEP-TH 0305248;%%



\del
\bibitem{Manin:1988??}%
Y.~I. Manin, \T{Quantum groups and non-commutative geometry,} Centre de
  Recherches Mathematiques, Universite de Montreal (1988).

\bibitem{Bouwknegt:1996mn}
  P.~Bouwknegt, J.~G.~McCarthy and P.~van Nieuwenhuizen,
  \T{Fusing the coordinates of quantum superspace,}
  Phys.\ Lett.\  B {\bf 394} (1997) 82
  [arXiv:hep-th/9611067].
  %%CITATION = PHLTA,B394,82;%%
\enddel



\bibitem{Klemm:2001yu}%
  D.~Klemm, S.~Penati and L.~Tamassia,
  \T{Non(anti)commutative superspace,}		\cut
  Class.\ Quant.\ Grav.\  {\bf 20} (2003) 2905	\endcut
  [arXiv:hep-th/0104190].
  %%CITATION = CQGRD,20,2905;%%



\bibitem{Ferrara:2000mm}%
  S.~Ferrara, M.~A.~Lledo,
  \T{Some aspects of deformations of supersymmetric field theories,}\cut
  JHEP {\bf 0005} (2000) 008					\endcut
  [arXiv:hep-th/0002084].
  %%CITATION = JHEPA,0005,008;%%


\bibitem{Kosinski:2000xu}%
  P.~Kosinski, J.~Lukierski and P.~Maslanka,
  \T{Quantum deformations of space-time SUSY and noncommutative superfield 
	theory,}
  [arXiv:hep-th/0011053].
  %%CITATION = HEP-TH/0011053;%%

%\cite{Ooguri:2003qp}
\bibitem{Ooguri:2003qp}%
H.~Ooguri and C.~Vafa,
\T{The C-deformation of gluino and non-planar diagrams,}
Adv.\ Theor.\ Math.\ Phys.\  {\bf 7} (2003) 53
~[arXiv:hep-th/0302109].
%%CITATION = HEP-TH 0302109;%%

\del
%\cite{Ooguri:2003tt}
\bibitem{Ooguri:2003tt}%
H.~Ooguri and C.~Vafa,
\T{Gravity induced C-deformation,}
~[arXiv:hep-th/0303063].
%%CITATION = HEP-TH 0303063;%%

%\cite{Grassi:2001ug}
\bibitem{Grassi:2001ug}
P.~A.~Grassi, G.~Policastro, M.~Porrati and P.~Van Nieuwenhuizen,
\T{Covariant quantization of superstrings without pure spinor constraints,}
JHEP {\bf 0210} (2002) 054
~[arXiv:hep-th/0112162].
%%CITATION = HEP-TH 0112162;%%

%\cite{Berkovits:2003kj}
\bibitem{Berkovits:2003kj}
N.~Berkovits and N.~Seiberg,
\T{Superstrings in graviphoton background and N = 1/2 + 3/2 supersymmetry,}
JHEP {\bf 0307} (2003) 010
~[arXiv:hep-th/0306226].
%%CITATION = HEP-TH 0306226;%%

\enddel



%\cite{deBoer:2003dn}
\bibitem{deBoer:2003dn}%
J.~de Boer, P.~A.~Grassi and P.~van Nieuwenhuizen,
\T{Non-commutative superspace from string theory,}
Phys.\ Lett.\ B {\bf 574} (2003) 98
~[arXiv:hep-th/0302078].
%%CITATION = HEP-TH 0302078;%%


	

\bibitem{Berkovits:1994wr}%
  N.~Berkovits,
  \T{Covariant quantization of the Green-Schwarz superstring in a Calabi-Yau 
	background,}
  Nucl.\ Phys.\  B {\bf 431} (1994) 258
  [arXiv:hep-th/9404162].
  %%CITATION = NUPHA,B431,258;%%

\bibitem{Grassi:2003cm}%
  P.~A.~Grassi, G.~Policastro and P.~van Nieuwenhuizen,
  \T{An introduction to the covariant quantization of superstrings,}
  Class.\ Quant.\ Grav.\  {\bf 20} (2003) S395
  [arXiv:hep-th/0302147].
  %%CITATION = CQGRD,20,S395;%%



%\cite{Berkovits:2002zk}
\bibitem{Berkovits:2002zk}
N.~Berkovits,
\T{ICTP lectures on covariant quantization of the superstring,}
~[arXiv:hep-th/0209059].
%%CITATION = HEP-TH 0209059;%%


\bibitem{Tyutin:2001iz}%
  I.~V.~Tyutin,
  \T{The general form of the star-product on the Grassman algebra,}%\cut
  Theor.\ Math.\ Phys.\  {\bf 127} (2001) 619
  [Teor.\ Mat.\ Fiz.\  {\bf 127} (2001) 253]			%\endcut
  [arXiv:hep-th/0101046].
  %%CITATION = TMFZA,127,253;%%




\bibitem{Chepelev:2003ga}%		% 31
  I.~Chepelev and C.~Ciocarlie,
  \T{A path integral approach to noncommutative superspace,}	\cut
  JHEP {\bf 0306} (2003) 031					\endcut
  [arXiv:hep-th/0304118].
  %%CITATION = JHEPA,0306,031;%%



%\cite{Berkovits:2000fe}
\bibitem{Berkovits:2000fe}%
N.~Berkovits,
\T{Super-Poincar\'e covariant quantization of the superstring,}%\cut
JHEP {\bf 0004} (2000) 018				%	\endcut
~[arXiv:hep-th/0001035].
%%CITATION = HEP-TH 0001035;%%




\bibitem{Cornalba:2002cu}
  L.~Cornalba, M.~S.~Costa and R.~Schiappa,
  \T{D-brane dynamics in constant Ramond-Ramond potentials and
  noncommutative geometry,}
  Adv.\ Theor.\ Math.\ Phys.\  {\bf 9} (2005) 355
  [arXiv:hep-th/0209164].
  %%CITATION = 00203,9,355;%%



\bibitem{Berkovits:2002ag}%
  N.~Berkovits and V.~Pershin,
  \T{Supersymmetric Born-Infeld from the pure spinor formalism of the open 
	superstring,}
  JHEP {\bf 0301} (2003) 023
  [arXiv:hep-th/0205154].
  %%CITATION = JHEPA,0301,023;%%


\bibitem{Berkovits:2001ue}%
  N.~Berkovits and P.~S.~Howe,
  \T{Ten-dimensional supergravity constraints from the pure spinor 
	formalism for the superstring,}
  Nucl.\ Phys.\  B {\bf 635} (2002) 75
  [arXiv:hep-th/0112160].
  %%CITATION = NUPHA,B635,75;%%



\bibitem{Bedoya:2006ic}%
  O.~A.~Bedoya and O.~Chandia,
  \T{One-loop conformal invariance of the type II pure spinor superstring 
	in a curved background,}
  JHEP {\bf 0701} (2007) 042
  [arXiv:hep-th/0609161].
  %%CITATION = JHEPA,0701,042;%%
				
\bibitem{Guttenberg:2007th}%
  S.~Guttenberg,
  \T{Superstrings in General Backgrounds,} 
		~ [http://www.ub.tuwien.ac.at/diss/AC05035309.pdf]
	\\PhD-thesis 2007, TU-Vienna.
			


\end{thebibliography}
\end{document}

\del

%\cite{Bachas:2000ik}
\bibitem{Bachas:2000ik}
C.~Bachas, M.~R.~Douglas and C.~Schweigert,
\T{Flux stabilization of D-branes,}
JHEP {\bf 0005} (2000) 048
~[arXiv:hep-th/0003037].
%%CITATION = HEP-TH 0003037;%%

%\cite{Alekseev:2000fd}
\bibitem{Alekseev:2000fd}
A.~Y.~Alekseev, A.~Recknagel and V.~Schomerus,
\T{Brane dynamics in background fluxes and non-commutative geometry,}
JHEP {\bf 0005} (2000) 010
~[arXiv:hep-th/0003187].
%%CITATION = HEP-TH 0003187;%%

%\cite{Kling:2000dy}
\bibitem{Kling:2000dy}
A.~Kling, M.~Kreuzer and J.~G.~Zhou,
\T{SU(2) WZW D-branes and quantized worldvolume U(1) flux on S(2),}
Mod.\ Phys.\ Lett.\ A {\bf 15} (2000) 2069
~[arXiv:hep-th/0005148].
%%CITATION = HEP-TH 0005148;%%

%\cite{Takayanagi:2001gu}
\bibitem{Takayanagi:2001gu}
T.~Takayanagi and T.~Uesugi,
\T{Flux stabilization of D-branes in NSNS Melvin background,}
Phys.\ Lett.\ B {\bf 528} (2002) 156
~[arXiv:hep-th/0112199].
%%CITATION = HEP-TH 0112199;%%

%% \cite{Callan:1986bc}
% \bibitem{Callan:1986bc}
%  C.~G.~.~Callan, C.~Lovelace, C.~R.~Nappi and S.~A.~Yost,
%  \T{String Loop Corrections To Beta Functions,}
%  Nucl.\ Phys.\  B {\bf 288} (1987) 525.
%  %%CITATION = NUPHA,B288,525;%%

%\bibitem{Berkovits:2003kj}
%  N.~Berkovits and N.~Seiberg,
%  \T{Superstrings in graviphoton background and N = 1/2 + 3/2 supersymmetry,}
%  JHEP {\bf 0307} (2003) 010
%  [arXiv:hep-th/0306226].
%  %%CITATION = JHEPA,0307,010;%%

\bibitem{Siegel:1985xj}
  W.~Siegel,
  \T{Classical Superstring Mechanics,}
  Nucl.\ Phys.\  B {\bf 263} (1986) 93.
  %%CITATION = NUPHA,B263,93;%%

\bibitem{Guttenberg:2007th}
  S.~Guttenberg,
  \T{Superstrings in General Backgrounds,} PhD-thesis 2007, TU-Vienna, 
	[http://www.ub.tuwien.ac.at/diss/AC05035309.pdf].